%% file: main.tex
\newif\ifarxiv
\arxivtrue

\ifarxiv
\documentclass[english,onecolumn,notitlepage,12pt,nofootinbib,amsmath,amssymb,superscriptaddress,aps,prd,longbibliography]{revtex4-1}
\usepackage{graphicx}
\usepackage[hidelinks]{hyperref}
\usepackage{multirow}

\usepackage[english]{babel}
\usepackage{newtxtext}

\else
\documentclass[entropy,article,submit,moreauthors,pdftex]{Definitions/mdpi} 
\fi

\usepackage{amsfonts}
\usepackage{amsmath}
\usepackage{amsthm}
\usepackage{url}
\usepackage[capitalize]{cleveref}

\DeclareMathOperator*{\argmax}{arg\,max}

\newcommand{\dd}{{\;d}}

\newcommand{\DKL}{D_\mathrm{KL}}
\newcommand{\II}{\mathbf{I}}

\newcommand{\M}{M}
\newcommand{\IXM}{I(X;\M)}
\newcommand{\IYM}{I(Y;\M)}
\newcommand{\IYX}{I(X;Y)}
\newcommand{\LIB}{\mathcal{L}_\mathrm{IB}}
\newcommand{\LIBup}{\hat{\mathcal{L}}_\mathrm{IB}}

\newcommand{\LsqIB}{\mathcal{L}_\mathrm{sqIB}}
\newcommand{\LsqIBup}{\hat{\mathcal{L}}_\mathrm{sqIB}}

\newcommand{\MIestBase}{\hat{I}_{\theta}}
\newcommand{\MIest}{\MIestBase(X;\M)}

\newcommand{\VM}{R}

\newcommand{\CeLoss}{\mathbb{E}_{Q_\theta(Y,M)}\big[\log P_\phi(Y|\M) \big]}

\newcommand{\figsfx}{v1-256-p50}

\newcommand{\myabstract}{Information bottleneck (IB) is a technique for extracting information in one random variable $X$ that is relevant for predicting another random variable $Y$.
IB works by encoding $X$ in a compressed ``bottleneck'' random variable $\M$ from which $Y$ can be accurately decoded. However, finding the optimal bottleneck variable  
involves a difficult optimization problem, which until recently has been considered for only two limited cases: discrete $X$ and $Y$ with small state spaces, and continuous $X$ and $Y$ with a Gaussian joint distribution (in which case optimal encoding and decoding maps are linear).  We propose a method for performing IB on arbitrarily-distributed discrete and/or continuous $X$ and $Y$, while allowing for nonlinear encoding and decoding maps. Our approach relies on a novel non-parametric  upper bound for  mutual information.  We describe how to implement our method using neural networks. We then show that it achieves better performance than the recently-proposed ``variational IB'' method on several real-world datasets.}

\ifarxiv

\begin{document}

\title{Nonlinear Information Bottleneck}

\author{Artemy Kolchinsky}
\email[]{artemyk@gmail.com}
\affiliation{\small Santa Fe Institute, 1399 Hyde Park Road, Santa Fe, NM, 87501, USA}
\author{Brendan D. Tracey}
\affiliation{\small Santa Fe Institute, 1399 Hyde Park Road, Santa Fe, NM, 87501, USA}
\affiliation{\small Dept Aeronautics \& Astronautics, Massachusetts Institute of Technology, Cambridge, MA 02139, USA}
\author{David H. Wolpert}
\affiliation{\small Santa Fe Institute, 1399 Hyde Park Road, Santa Fe, NM, 87501, USA}
\affiliation{\small Complexity Science Hub, Vienna, Austria}
\affiliation{\small Arizona State University, Tempe, AZ 85287, USA}

\begin{abstract}
\myabstract
\end{abstract}

\maketitle

\else

\firstpage{1} 
\makeatletter 
\setcounter{page}{\@firstpage} 
\makeatother
\pubvolume{xx}
\issuenum{1}
\articlenumber{5}
\pubyear{2019}
\copyrightyear{2019}
\history{Received: 16 October 2019; Accepted: 28 November 2019; Published: date}





\Title{Nonlinear Information Bottleneck}

\newcommand{\orcidauthorA}{0000-0002-3518-9208} 
\newcommand{\orcidauthorB}{0000-0003-3105-2869} 

\Author{ {Artemy Kolchinsky} $^{1,*}$\orcidA{}, Brendan D. Tracey $^{1,2}$ and David H. Wolpert $^{1,3,4}$\orcidB{}}

\AuthorNames{Firstname Lastname, Firstname Lastname and Firstname Lastname}

\address{%
$^{1}$ \quad Santa Fe Institute, 1399 Hyde Park Road, Santa Fe, NM 87501, USA; tracey.brendan@gmail.com(B.D.T.); david.h.wolpert@gmail.com(D.H.W.)\\
$^{2}$ \quad Department of Aeronautics \& Astronautics, Massachusetts Institute of Technology, Cambridge, MA 02139, USA\\
$^{3}$ \quad Complexity Science Hub, Vienna 1080, Austria\\
$^{4}$ \quad {Center for Bio-social Complex Systems}, Arizona State University, Tempe, AZ 85281, USA
}

\corres{Correspondence: artemyk@gmail.com}





\abstract{\myabstract
}








\begin{document}


\fi

\section{Introduction}

Imagine that one has two random variables, an ``input'' random variable $X$ and an ``output'' random variable $Y$, and that one wishes to use $X$ to predict $Y$.  
In some situations, it is useful to extract a compressed representation of  $X$ that is relevant for predicting $Y$. 
%
This problem is formally considered by the \emph{information bottleneck} (IB) method \cite{tishby_information_1999,dimitrov_neural_2001,samengo_information_2002}.  
IB proposes to find a ``bottleneck'' variable $\M$ which 
maximizes prediction, formulated in terms of the mutual information  $\IYM$, given a constraint on compression, formulated in terms of the mutual information $\IXM$. Formally, this can be stated in terms of the  constrained optimization problem
\begin{equation}
\argmax_{\M \in \Delta} \IYM \quad\text{s.t.}\quad \IXM \le R  \,,
\label{eq:constrainedproblem}
\end{equation}
where $\Delta$ is the set of random variables $\M$ that obey  the
Markov condition $Y-X-\M$~\citep{witsenhausen_conditional_1975,ahlswede_source_1975,goos_information_2003}.   This Markov condition states that $\M$ is conditionally independent of $Y$ given $X$, and it guarantees that any information that $\M$ has about $Y$ is extracted from $X$.  
The maximal  value of $\IYM$ for each possible compression value $R$ forms what is called the \emph{IB curve}~\cite{tishby_information_1999}.



%
The following example illustrates how IB might be used. 
Suppose that a remote weather station makes detailed recordings of meteorological data ($X$), which are then encoded and sent to a central server ($\M$) and used to predict weather conditions for the next day ($Y$).  If the channel between the weather station and server has low capacity, then the information transmitted from the weather station to the server must be compressed.  Minimizing the IB objective amounts to finding a compressed representation of meteorological data which can be transmitted across a low capacity channel (have low $\IXM$) and used to optimally predict future weather (have high $\IYM$).
The IB curve specifies the trade-off between channel capacity and  accurate prediction.

Numerous applications of IB exist in domains such as clustering~\citep{slonim2000document,tishby_data_2001}, coding theory and quantization~\citep{cardinal2003compression,zeitler2008design,courtade2011multiterminal,lazebnik2008supervised}, speech and image recognition~\cite{winn2005object,hecht2009speaker,yaman2012bottleneck,van2017information,van2019speech}, 
and cognitive science~\citep{zaslavsky2018efficient}.   
Several recent papers have also drawn connections between IB
and {supervised learning}, in particular, classification using neural networks~\cite{rodriguez_galvez_information_2019,hafez2019information}. 
In this context, $X$ typically represents input vectors, $Y$ the output classes, and $\M$ the intermediate representations used by the network, such as the activity of hidden layer(s)~\citep{tishby2015deep}. 
Existing research has considered whether intermediate representations that are optimal in the IB sense (i.e., close to the IB curve) may be better in terms of 
generalization error~\citep{shamir2010learning,tishby2015deep,vera2018role}, robustness to adversarial inputs~\citep{alemi_deep_2016}, detection of out-of-distribution data~\citep{alemi2018uncertainty}, or provide more  ``interesting'' or ``useful'' intermediate representations of inputs~\citep{amjad2018not}.
Other related research has investigated whether stochastic gradient descent (SGD) training dynamics may drive hidden layer representations towards IB optimality~\cite{shwartz2017opening,saxe2018information}.

In practice, optimal bottleneck variables are usually not found 
by solving the constrained optimization problem of Equation \eqref{eq:constrainedproblem}, but rather 
by finding $\M$ that maximize
the so-called \emph{IB Lagrangian}
~\citep{tishby_information_1999,shamir2010learning,goos_information_2003}, 
\begin{equation}
\LIB(\M) := \IYM - \beta \IXM .
\label{eq:unconstrainedproblem}
\end{equation}
$\LIB$ is the Lagrangian relaxation~\cite{lemarechal2001lagrangian} of the constrained optimization problem of Equation \eqref{eq:constrainedproblem}, and $\beta$ is a Lagrange multiplier that enforces the constraint $\IXM \le R$. 
  In practice, $\beta \in [0,1]$ serves as a  parameter that controls the trade-off between compression
and prediction.
As $\beta \to 1$, IB will favor maximal compression of $X$; for $\beta = 1$ (or any $\beta \ge 1$) the optimal   $\M$  will satisfy $\IXM =\IYM =0$.  As $\beta \to 0$, IB will favor prediction of $Y$; for $\beta = 0$ (or any $\beta \le 0$), there is no penalty on $\IXM$ and the optimal $\M$ will satisfy $\IYM = \IYX$, the maximum possible. 
It is typically easier to optimize $\LIB$ than Equation \eqref{eq:constrainedproblem}, since the latter involves a complicated non-linear constraint. For this reason, optimizing $\LIB$ has become standard in the IB literature~\cite{tishby_information_1999,goos_information_2003,rodriguez_galvez_information_2019,hafez2019information,chechik_information_2005,alemi_deep_2016,shamir2010learning}.


However,  in recent work~\cite{kolchinsky2018caveats}  we showed that whenever $Y$ is a deterministic function of $X$ (or close to being one), optimizing $\LIB$ is not longer equivalent to optimizing Equation \eqref{eq:constrainedproblem}. 
In fact, when $Y$ is a deterministic function of $X$,  the same $\M$ will optimize $\LIB$ for all values of $\beta$, meaning that the IB curve cannot be explored by optimizing $\LIB$ while sweeping $\beta$.   
This is a serious issue in supervised learning scenarios (as well as some other domains), where it is very common for the output $Y$ to be a deterministic function of the input $X$.  
Nonetheless, the IB curve can still be explored by optimizing the following simple modification of the IB Lagrangian, which we called the \emph{squared-IB Lagrangian}~\cite{kolchinsky2018caveats},
\begin{equation}
\LsqIB(\M) := \IYM - \beta \IXM^2 
\label{eq:sqiblagrangian}
\end{equation}
where $\beta \ge 0$ is again a parameter that controls the trade-off between compression 
and prediction. 
Unlike the case for $\LIB$, there is always a one-to-one correspondence between $\M$ that optimize $\LsqIB$ and solutions to Equation \eqref{eq:constrainedproblem}, regardless of the relationship between $X$ and $Y$.  
In the language of optimization theory, the squared-IB Lagrangian is a ``scalarization'' of the multi-objective problem $\{ \min \IXM, \max \IYM \}$~\cite{miettinen_nonlinear_1998}. Importantly, unlike $\LIB$, there can be non-trivial optimizers of $\LsqIB$ even for $\beta \ge 1$; the relationship between $\beta$ and corresponding solutions on the IB curve has been analyzed in~\cite{anonymous2020the}. 
In that work, it was also shown that the objective function of \cref{eq:sqiblagrangian} is part of a general family of objectives $\IYM - \beta F(\IXM)$, where $F$ is any monotonically-increasing and strictly convex function, 
all of which can be used to explore the IB curve. 


Unfortunately, optimizing the IB Lagrangian and squared-IB Lagrangian remains a difficult problem. 
First, both objectives are non-convex, so there is no guarantee that a global optimum can be found.  Second, finding even a local optimum  requires evaluating the mutual information terms $\IXM$ and $\IYM$, which can involve  intractable integrals.  
For this reason, until recently IB has been mainly developed for two limited cases. The first case is where
$X$ and $Y$ are discrete-valued and have a small number of possible outcomes~\cite{tishby_information_1999}.  There, one can  
explicitly represent the full   
 \emph{encoding map} (the condition probability distribution of $\M$ given $X$) 
 during optimization, and the relevant integrals become tractable finite sums.
The second case is when $X$ and $Y$ are continuous-valued and jointly Gaussian.  Here, the IB optimization problem can be solved analytically, and the resulting 
encoding and decoding maps 
are linear~\cite{chechik_information_2005}. 

In this work, we propose a method for performing IB in much more general settings, which we call \emph{nonlinear information bottleneck}, or \emph{nonlinear IB} for short.  Our method assumes that $M$ is a continuous-valued random variable, but $X$ and $Y$ can be either discrete-valued (possibly with many states) or continuous-valued, and with any desired joint distribution.  Furthermore, as suggested by the term nonlinear IB, the encoding and decoding maps can be nonlinear.

To carry out nonlinear IB, we derive a lower bound on $\LIB$ (or, where appropriate, $\LsqIB$) which can be maximized using gradient-based methods.
As we describe in the next section, our approach makes use of the following techniques:
\begin{itemize}
	\item We represent the distribution over $X$ and $Y$ using a finite number of data samples.
	\item We represent the encoding map $p(m\vert x)$ and the \emph{decoding map} $p(y\vert m)$ as parameterized conditional distributions.  
	\item We use a variational lower bound for the prediction term $\IYM$, and 
	non-parametric upper bound for the compression term $\IXM$, which we developed in earlier work~\cite{kolchinsky_upperbound_2017}.
\end{itemize}

Note that three recent papers have suggested other ways of optimizing the IB Lagrangian  
in general settings~\cite{chalk_relevant_2016,alemi_deep_2016,achille2016information}. 
These papers use  variational upper bounds on the compression term $\IXM$, which is different from our non-parametric upper bound. 
A  detailed comparison is provided in \cref{sec:Relation-to-Existing}. In that section, we also relate our approach to other work in machine learning. 

In \cref{sec:experiments}, we explain how to implement our approach using standard neural network techniques. We demonstrate its performance on several real-world datasets, and compare it to the recently-proposed  \emph{variational IB} method~\cite{alemi_deep_2016}.

\section{Proposed Approach}

\label{subsec:Overview-of-approach}

In the following, we use $H(\cdot)$ for Shannon entropy, $I(\cdot ; \cdot)$ for mutual information [MI],  $\DKL(\cdot \Vert \cdot)$ for Kullback--Leibler [KL] divergence.  
All information-theoretic quantities are in units of bits, and all $\log$s are base-2. We use $\mathcal{N}(\mu, \Sigma)$ to indicate the probability density function of a multivariate Gaussian  with mean $\mu$ and covariance matrix $\Sigma$. 
We use notation like $\mathbb{E}_{P(X)}[f(X)] = \int P(x) f(x) \, dx$ to indicate expectations, where $f(x)$ is some function and $P(x)$ some probability distribution. We use $\delta(\cdot, \cdot)$ for the Kronecker delta.

Let the input random variable $X$ and the output random variable $Y$ be distributed according to some joint distribution $Q(x,y)$, with marginals indicated by $Q(y)$ and $Q(x)$. We assume that we are provided with a ``training dataset'' $\mathcal{D} = \{(x_1, y_1), \dots, (x_N, y_N)\}$, which contains $N$ input--output pairs sampled IID from $Q(x,y)$. Let $\M$ indicate the bottleneck random variable, with outcomes in $\mathbb{R}^d$.  In the derivations in this section, we assume that $X$ and $Y$ are continuous-valued, but our approach extends immediately to the discrete case (with some integrals replaced by sums).

Let the conditional probability $P_\theta(m|x)$ indicate a parameterized \emph{encoding map} from input $X$ to the bottleneck variable $\M$, where $\theta$ is a vector of parameters. 
Given an encoding map, one can compute the MI between $X$ and $\M$, $I_\theta(X;\M)$, using the joint distribution $Q_\theta(x,m) := P_\theta(m|x)Q(x)$. Similarly, one can compute the MI between $Y$ and $\M$, $I_\theta(Y;\M)$, using the joint distribution 
\begin{align}
Q_\theta(y,m) := \int P_\theta(m|x)Q(x,y) \dd x \,.
\label{eq:integral-2}
\end{align}

We now consider the IB Lagrangian, Equation \eqref{eq:unconstrainedproblem}, as a function of the encoding map parameters, 
\begin{align}
\LIB(\theta) & :=  I_\theta(Y;M) - \beta I_\theta(X;M)  \,.
\label{eq:overview-LIB}
\end{align}
In this parametric setting, we seek parameter values that maximize $\LIB(\theta)$.   Unfortunately,  this optimization problem is usually intractable due to the difficulty of computing the integrals in  Equation \eqref{eq:integral-2} and in the MI terms of Equation \eqref{eq:overview-LIB}. Nonetheless, it is possible to carry out an approximate form of IB by maximizing a tractable lower bound on $\LIB$, which we now derive.

First, consider any conditional probability $P_\phi(y|m)$ of outputs given bottleneck variable, where $\phi$ is a vector of parameters, which we call  the \emph{(variational) decoding map}.
Given $P_\phi(y|m)$,  the non-negativity of KL divergence leads to the following {variational} lower bound on the first MI term in Equation \eqref{eq:overview-LIB},
\begin{align}
I_\theta(Y;M) & = H(Q(Y)) - H(Q_\theta(Y|M)) \nonumber \\
       & \ge H(Q(Y)) - H(Q_\theta(Y|M)) - \DKL(Q_\theta(Y|M)\Vert P_\phi(Y|M)) 
\nonumber \\ 
& = H(Q(Y)) + \CeLoss  \,,
\label{eq:decoder-bound}
\end{align}
where in the last line we've used the following identity,
\begin{align}
    -\CeLoss =\DKL( Q_\theta(Y|\M)\Vert P_{\phi}(Y|\M)) + H(Q_\theta(Y|\M)). \label{eq:rel0}
\end{align}
Note that the inequality of Equation \eqref{eq:decoder-bound} holds for any choice of $P_\phi(y|m)$, and becomes an equality when $P_\phi(y|m)$ is equal to the ``optimal'' decoding map $Q_\theta(y|m)$ (as would be computed from Equation \eqref{eq:integral-2}).  Moreover,  
the bound becomes tighter as the KL divergence between $P_\phi(y|m)$ and 
 $Q_\theta(y|m)$ gets smaller. Below, we will maximize the RHS of Equation \eqref{eq:decoder-bound} with respect to $\phi$, thereby  bringing  $P_\phi(y|m)$ closer to $Q_\theta(y|m)$. 


%
%
%
%
%

%

It remains to upper bound the $I_\theta(X;\M)$ term in Equation \eqref{eq:overview-LIB}. 
To proceed, we first approximate the joint distribution of $X$ and $Y$ with the empirical distribution in the training dataset, 
\begin{align}
Q(x,y)\approx \frac{1}{N}\sum_i \delta(x_i,x)\delta(y_i,y).
\label{eq:emp}
\end{align} 
We then assume 
that the encoding map is the sum of a deterministic function $f_\theta(x)$ plus Gaussian noise, 
\begin{align}
M = f_\theta(X) + Z,
%
\label{eq:m-as-func-of-x}
\end{align}
where $(Z \vert X=x) \sim \mathcal{N}(f_\theta(x), \Sigma_{\theta}(x))$. 
Note that 
the  noise covariance $\Sigma_{\theta}(x)$ can depend both on the parameters $\theta$ and the outcome of $X$ (i.e., the noise can be {heteroscedastic}).  
Combining   Equation \eqref{eq:emp} and Equation \eqref{eq:m-as-func-of-x} implies  that the bottleneck variable $\M$ will be distributed as a mixture of $N$ equally-weighted Gaussian components, with component $i$ having distribution $\mathcal{N}(f_\theta(x_i), \Sigma_{\theta}(x_i))$.  We can then employ the following non-parametric upper bound on MI, which was derived in a recent paper~\cite{kolchinsky_upperbound_2017}:
\begin{align}
\label{eq:MIestA}
I_\theta(X;M) \le \MIest := -\frac{1}{N} \sum_i \log \frac{1}{N}\sum_j e^{-\DKL\big[ \mathcal{N}(f_\theta(x_i), \Sigma_{\theta}(x_i)) \big\Vert \mathcal{N}(f_\theta(x_j), \Sigma_{\theta}(x_j))\big]} .
\end{align}
(Note that the published version of \cite{kolchinsky_upperbound_2017} contains some typos which are corrected in the latest arXiv version at \href{http://arxiv.org/abs/1706.02419}{arxiv.org/abs/1706.02419}.)

Equation \eqref{eq:MIestA} bounds the MI in terms of the pairwise KL divergences between the Gaussian components of  the mixture distribution of $\M$.  It is useful because the KL divergence between two $d$-dimensional Gaussians has a closed-form expression, 
\begin{align}
\DKL\big[ \mathcal{N}(\mu',\Sigma') \big\Vert \mathcal{N}(\mu,\Sigma)\big] = \frac{1}{2}\left[ \ln \frac{\det \Sigma}{\det \Sigma'} + (\mu' - \mu)\Sigma^{-1}(\mu' - \mu) + \mathrm{tr}(\Sigma^{-1} \Sigma') - d \right] .
\label{eq:GaussianKL}
\end{align}
Furthermore,  in the special case when all components have the same covariance and can be grouped into well-separated clusters, the upper bound of Equation \eqref{eq:MIestA} becomes tight~\cite{kolchinsky_upperbound_2017}.  As we will see below, this special case is a commonly encountered solution to the optimization problem considered here.

Combining Equation \eqref{eq:decoder-bound} and Equation \eqref{eq:MIestA} provides the following tractable lower bound for the IB Lagrangian,
\begin{align}
\LIB(\theta) & \ge \LIBup(\theta,\phi) :=  \CeLoss - \beta \MIest 
\label{eq:lbA}
\end{align}
where we dropped the additive constant $H(Q(Y))$ (which does not depend on the parameter values and is therefore irrelevant for optimization).  
We refer to Equation \eqref{eq:lbA} as the \emph{nonlinear IB} objective.

As mentioned in the introduction, in cases where $Y$ is a deterministic function of $X$ (or close to being one), it is no longer possible to explore the IB curve by optimizing the IB Lagrangian for different values of $\beta$~\cite{kolchinsky2018caveats,anonymous2020the,rodriguez_galvez_information_2019}.  Nonetheless, it is always possible to explore the IB curve by instead optimizing the squared-IB Lagrangian, Equation \eqref{eq:sqiblagrangian}.  The above derivations also lead to the following tractable lower bound for the squared-IB Lagrangian,
\begin{align}
\LsqIB(\theta)  & \ge \LsqIBup(\theta,\phi) :=   \CeLoss - \beta \big[\MIest\big]^2 .
\label{eq:lbB}
\end{align}

Note that maximizing the expectation term $\CeLoss$ is equivalent to minimizing the usual cross-entropy loss in supervised learning. (Note that {mean squared error,} the typical  loss function used for training regression models, can also be interpreted as a cross-entropy term~\cite[p.132--134]{goodfellow2016deep}.). From this point of view, Equation \eqref{eq:lbA} and Equation \eqref{eq:lbB} can be interpreted as adding an information-theoretic regularization term to the regular objective of supervised learning.  

For optimization purposes, the compression term $\MIest$ can be computed from data using Equations \eqref{eq:MIestA} and \eqref{eq:GaussianKL}, 
 while the expectation term $\CeLoss$ can be estimated as $\CeLoss \approx \frac{1}{N} \sum_i \log P_{\phi}(y_i|m_i)$, 
where $m_i$  indicates samples from  $P_\theta(m|x_i)$. 
Assuming that $f_\theta$ is differentiable with respect to $\theta$  and $P_\phi$ is differentiable with respect to $\phi$, the optimal  $\theta$ and $\phi$ can be selected by using gradient-based methods to maximize Equation \eqref{eq:lbA} or Equation \eqref{eq:lbB}, as desired.   In practice, this optimization will typically be done using stochastic gradient descent (SGD), i.e., by  computing the gradient using randomly sampled mini-batches rather than the whole training dataset.  In fact, mini-batching becomes necessary for large datasets, since evaluating $\MIest$ involves $O(n^2)$ operations, where $n$ is the number of data points in the batch used to compute the gradient, which becomes prohibitively slow for very large $n$.  
At the same time, $\MIest$ is closely related to kernel-density estimators~\cite{kolchinsky_upperbound_2017}, and it is known that the number of samples required for accurate kernel-density estimates grows rapidly as dimensionality increases~\cite{silverman2018density}.  Thus,  mini-batches should not be too small when $d$ (the dimensionality of the bottleneck variable) is large.  In some cases, it may be useful to estimate the gradient of  $\CeLoss$ and the gradient of $\MIest$ using mini-batches of different sizes. 
More implementation details are discussed below in \cref{sec:implementation}.

Note that the approach described here is somewhat different (and simpler) than in previous versions of this manuscript~\cite{kolchinsky-slip-rnn-2016,nonlinearIBv1}. In previous versions, we represented the marginal distribution $Q(x)$ with a mixture of Gaussians, rather than with the empirical distribution in the training data. However, we found that this increased complexity but was not necessary for good performance. Furthermore, we previously focused only on optimizing a bound on the IB Lagrangian, Equation \eqref{eq:lbA}. In subsequent work~\cite{kolchinsky2018caveats}, we showed that the IB Lagrangian is inadequate for many supervised learning scenarios, including some of those explored in \cref{sec:Results}, and that the squared-IB Lagrangian should be used instead. In this work, we report performance when optimizing Equation \eqref{eq:lbB}, a bound on the squared-IB Lagrangian.


%

\section{Relation to Prior Work}
\label{sec:Relation-to-Existing}

In this section, we relate our proposed method to prior work in machine learning.

\subsection{Variational IB}
\label{subsec:variationalib}

Recently, there have been three other proposals for performing IB for continuous and possibly non-Gaussian random variables using neural networks~\cite{chalk_relevant_2016,alemi_deep_2016,achille2016information}, the most popular of which is  called \emph{variational IB} (VIB)~\cite{alemi_deep_2016}. 
As in our approach, these methods propose tractable lower bounds on the $\LIB$ objective. They employ the same variational bound for the prediction MI term $\IYM$ as our Equation \eqref{eq:decoder-bound}. 
These methods differ from ours, however, in how they bound the compression term, $I_\theta(X;M)$. In particular, they all use some form of the following variational  upper bound, 
\begin{align}
\begin{split}
I_\theta(X;M) &= \DKL(P_\theta(\M|X)\Vert \VM(\M)) - \DKL(P_\theta(\M) \Vert \VM(M)) \le \DKL(P_\theta(\M|X)\Vert \VM(\M))  \,,
\end{split}
\label{eq:variational-xm-bound}
\end{align}
where  $\VM$ is some surrogate marginal distribution over the bottleneck variable $\M$. 
Combining with Equation \eqref{eq:decoder-bound} leads to the following variational lower bound for $\LIB$,
\begin{equation}
\LIB(\M) \ge
\CeLoss 
-  \beta \DKL(P_\theta(M|X)\Vert \VM(M)) + \text{const} \, .
\label{eq:variational-total-bound}
\end{equation}
The three aforementioned papers differ in how they define the surrogate marginal distribution $\VM$.  In~\cite{alemi_deep_2016}, $\VM$ is a standard multivariate normal distribution, $\mathcal{N}(0, \II)$. 
In~\cite{chalk_relevant_2016}, $\VM$ is a product of Student's \emph{t}-distributions. The scale and shape parameters of 
 each \emph{t}-distribution 
are optimized during training, in this way tightening the bound in Equation \eqref{eq:variational-xm-bound}. In~\cite{achille2016information}, two surrogate distributions are considered, the improper log-uniform and the log-normal, with the appropriate choice depending on the particular activation function (non-linearity) used in the neural network.


In addition, the encoding map $P_\theta(m|x)$ in~\cite{chalk_relevant_2016} and \cite{alemi_deep_2016} is a deterministic function plus Gaussian noise, same as in  Equation \eqref{eq:m-as-func-of-x}.  
In \cite{achille2016information}, the encoding map consists of a deterministic function with multiplicative, rather than additive, noise.



These alternative methods have potential advantages and disadvantages compared to our approach.  On one hand, they are more computationally efficient: Our non-parametric estimator of $\MIest$ requires $O(n^2)$ operations per mini-batch (where $n$ is the size of the mini-batch), while the variational bound of Equation \eqref{eq:variational-xm-bound} requires $O(n)$ operations.  On the other hand, our non-parametric estimator is expected to give a better estimate of the true MI $\IXM$~\cite{kolchinsky_upperbound_2017}. 
We provide a comparison between our approach and variational IB~\cite{alemi2018uncertainty} in  \cref{sec:Results}.

\subsection{Neural Networks and Kernel Density Entropy Estimates}

A key component of  our approach is using a differentiable upper bound on MI, $\MIest$. As discussed in \cite{kolchinsky_upperbound_2017}, this bound is related to non-parametric kernel-density  estimators of MI. 
See \cite{schraudolph_optimization_1995,schraudolph_gradient-based_2004,shwartz_fast_2005,torkkola2003feature,hlavavckova-schindler_causality_2007} for related work on using neural networks to optimize non-parametric estimates of information-theoretic functions.
This technique can also be related to kernel-based estimation of the likelihood of held-out data for neural networks (e.g.,~\cite{goodfellow2014generative}).  In these later approaches, however, the likelihood of held-out data is estimated only once, as a diagnostic measure once learning is complete. We  instead propose to train the network by directly incorporating our non-parametric estimator $\MIest$ in the objective function.

\subsection{Auto-Encoders}

Auto-encoders are unsupervised learning architectures that learn to reconstruct a copy of the input ${X}$, while using some intermediate representations (such as a hidden layer in a neural network).
Auto-encoders have some conceptual relationships to  IB, in  that the intermediate representations are sometimes restricted in terms of dimensionality, or with information-theoretic penalties on hidden layer coding length~\cite{hinton_autoencoders_1994,hinton_minimizing_1997}. Similar penalties have  also been explored in a supervised learning scenario in~\cite{deco_elimination_1993}.  In that work, however, hidden layer states were treated as discrete-valued, limiting the flexibility and information capacity of hidden representations.  

More recently, \emph{denoising auto-encoders}~\cite{vincent2008extracting} have attracted attention. Denoising auto-encoders constrain the amount of information passing from input to hidden layers by injecting noise into the hidden layer activity, similarly to our noisy mapping from the input to the bottleneck layer.  Previous work on auto-encoders has considered either penalizing hidden layer coding length \emph{or} injecting noise into the map, rather than combing the two as we do here. Moreover, denoising auto-encoders do not have a notion of an ``optimal'' noise level, since less noise will always improve prediction error on the training data. Thus, they  cannot directly adapt the noise level (as done in our method).

Finally, \emph{variational auto-encoders}~\cite{kingma2013auto} [VAEs] are recently-proposed architectures which learn generative models from unsupervised data (i.e., after training, they can be used to generate new samples that resemble training data).  Interestingly, the objective optimized in VAE training, called ``ELBO'',  contains both a prediction term and a compression term and can be seen as a special case of the variational IB objective~\cite{alemi_deep_2016,achille2016information,higgins2017beta,alemi2017fixing}.  In principle, it may be fruitful to replace the compression term in the ELBO with our MI estimator $\MIest$. Given our reported performance below, this may result in better compression, though it might also complicate sampling from the latent variable space. We leave this line of research for future work.

\section{Experiments}
\label{sec:experiments}

In this section, we first explain how to implement nonlinear IB using neural network techniques. We then  evaluate its on several datasets, and compare it to the variational IB (VIB) method. We demonstrate that, compared to VIB, nonlinear IB achieves better performance and uncovers different kinds of representations.

\subsection{Implementation}

\label{sec:implementation}


Any implementation of nonlinear IB requires a way to compute the encoding map $P_\theta(m|x)$ and decoding map $P_\phi(y|m)$, as well as a way to choose the parameters of these maps so as to maximize the nonlinear IB objective.  Here we explain how this can be done using standard neural network methods. 

The encoding map $P_\theta(m|x)$, Equation \eqref{eq:m-as-func-of-x}, is implemented in the following way: First, several neural network layers with parameters $\theta$ implement the (possibly nonlinear) deterministic function $f_\theta(x)$.  The output of these layers is then added to zero-centered Gaussian noise with covariance $\Sigma_{\theta}(x)$, which becomes the state of the \emph{bottleneck layer}. This is typically done via  the ``reparameterization trick''~\cite{kingma2013auto}, in which samples of Gaussian noise are passed through several deterministic layers (whose parameters are also indicated by $\theta$) and then added  to $f_\theta(x)$. 
Note that due to the presence of noise, the neural network is stochastic: even with parameters held constant, different states of the bottleneck layer are sampled during different NN evaluations. This stochasticity guarantees that the mutual information $\IXM$ is finite~\cite{saxe2018information,amjad2018not}.

In all of the experiments described below, the encoding map consists of two  layers with 128 ReLU neurons each, following by a layer of 5 linear neurons.
In addition, for simplicity we use a simple homoscedastic noise model: $\Sigma_{\theta}(x)= \sigma^2 \II$, where $\sigma^2$ is a parameter the sets the scale of the noise variance.  This noise model permits us to rewrite the MI bound of Equation \eqref{eq:MIestA} in terms of the following simple expression,
\begin{align}
\MIest = - \frac{1}{N}  \sum_{i} \log \frac{1}{N}  \sum_{j} e^{-\frac{1}{2\sigma^2}{\lVert f_\theta(x_i)-f_\theta(x_j) \rVert_2^2}}.
\end{align}
For purposes of comparison, we use this same homoscedastic noise model for both nonlinear IB and for VIB (note that the original VIB paper~\cite{alemi_deep_2016} used a  heteroscedastic noise model; investigating the performance of nonlinear IB with heteroscedastic noise remains for future work).  

In our runs, the noise parameter $\sigma^2$ was one of the trainable parameters in $\theta$.  The initial value of   $\sigma^2$  should be chosen with some care. If the initial $\sigma^2$ is too small,  the  Gaussian components that make up the mixture distribution of $\M$ will be many standard deviations away from each other and $\MIest$ (as well as $\IXM$) will be exponentially close to the constant $\log N$~\cite{kolchinsky_upperbound_2017}. In this case, the gradient of the compression term $\MIest$ with respect to $\theta$ will also be exponentially small, and the optimizer will not be able to learn to compress.  On the other hand, when $\sigma^2$ is too large, the resulting noise can swamp gradient information arising from the accuracy (cross-entropy) term, cause the optimizer to collapse to a ``trivial'' maximally-compressed model in which  $\IXM \approx \IYM \approx 0$.  Nonetheless, the optimization is robust to several orders of magnitude of variation of the initial value of $\sigma^2$. In the experiments below, we uses the initial value $\sigma^2 = 1$, which works sufficiently well in practice. ({Note that the scale of the noise can also be}
 trained by changing the parameters of the 5-neuron linear layer; thus, in our neural network architecture, having a trainable $\sigma^2$ is not strictly necessary.)


To implement the decoding map $P_\phi(y|m)$, the bottleneck layer states are passed through several deterministic neural network layers with parameters $\phi$.  
In the experiments described below, the decoding map is implemented with a single  layer with 128 ReLU neurons, followed by a linear output layer. 
The log decoding probability ($\log P_\phi(y|m)$) is then evaluated using the network output and an appropriately-chosen cost function: cross-entropy loss of the softmax of the output for classification, and mean squared error (MSE) of the output for regression.

In the experiments below, we use nonlinear IB to optimize the bound on the ``squared-IB Lagrangian'', Equation \eqref{eq:lbB}, rather than the bound on the IB Lagrangian, Equation \eqref{eq:lbA}.  For comparison purposes, we also optimize the following ``squared'' version of the VIB objective, Equation \eqref{eq:variational-total-bound},
\begin{equation}
{\mathcal{L}}_\text{sq-VIB} :=
\CeLoss 
-  \beta \big[\DKL(P_\theta(M|X)\Vert \VM(M))\big]^2 .
\label{eq:sq-variational-total-bound}
\end{equation}
As in the original VIB paper, we take $\VM(m)$ to be the standard Gaussian $\mathcal{N}(0,\II)$. 
We found that optimizing the squared-IB bounds, Equation \eqref{eq:lbB} and Equation \eqref{eq:sq-variational-total-bound}, produced quantitatively similar results to optimizing Equation \eqref{eq:lbA} and Equation \eqref{eq:variational-total-bound}, but was more numerically robust when exploring the full range of the IB curve. For an explanation of why this occurs, see the discussion and analysis in \cite{kolchinsky2018caveats}.  We report performance of nonlinear IB and VIB when optimizing bounds on the IB Lagrangian, Equation \eqref{eq:lbA} and Equation \eqref{eq:variational-total-bound}, in the Supplementary Material.

We use the Adam~\cite{kingma2015adam} optimizer with standard TensorFlow settings and mini-batches of size 256. To avoid over-fitting, we use early stopping: we split the training data into 80\% actual training data and 20\% validation data; training is stopped once the objective on the validation dataset did not improve for 50 epochs.

A TensorFlow implementation of our approach is provided at 
\texttt{\url{https://github.com/artemyk/nonlinearIB}}.  An independent PyTorch implementation is available at \texttt{\url{https://github.com/burklight/nonlinear-IB-PyTorch}}.

\subsection{Results}
\label{sec:Results}

We  report the performance of nonlinear IB on two different classification datasets (MNIST and FashionMNIST) and one regression dataset (California housing prices). We also  compare it with 
the recently-proposed variational IB (VIB) method~\cite{alemi_deep_2016}. Here we focus purely on the ability of these methods to 
optimize the IB objective on training and testing data. We leave for future work comparisons of these methods in terms of adversarial robustness~\cite{alemi_deep_2016}, detection of out-of-distribution data~\cite{alemi2018uncertainty}, and other desirable characteristics that may emerge from IB training.


We optimize both the nonlinear IB (Equation \eqref{eq:lbB}) and the VIB (Equation \eqref{eq:sq-variational-total-bound}) objectives for different values of $\beta$, producing a series of models that explore the trade-off between compression and prediction. 
We vary $\beta \in [10^{-3},2]$ for classification tasks and $\beta \in [10^{-5},2]$ for the regression task. These ranges were chosen empirically so that the resulting models fully explore the IB curve. 

To report our results, we use \emph{information plane} (info-plane) diagrams~\cite{shwartz2017opening}, which visualize the performance of different models in terms of the compression term  ($\IXM$, the x-axis) and the prediction term  ($\IYM$, the y-axis) both on training and testing data.   
For the info-plane diagrams, we use Monte Carlo sampling to get an accurate estimate of $\IXM$ terms. To estimate the  $\IYM =H(Y) - H(Y|M)$ term, we use two different approaches. For classification datasets, we approximate $H(Y)$ using the empirical entropy of the class labels in the dataset, and approximate the conditional entropy with the cross-entropy loss, $H(Y|\M) \approx -\CeLoss$. 
Note that the resulting MI estimate is an additive constant away from the cross-entropy loss. 
For the regression dataset, we approximate $H(Y)$ via the entropy of a Gaussian with variance $\text{Var}($Y$)$,  and approximate $H(Y|\M)$ via the entropy of a Gaussian with variance equal to the mean-squared-error.  This results in the estimate $\IYM \approx \frac{1}{2} \log (\text{Var}(Y) / \text{MSE})$.  
Finally, we also use scatter plots to visualize the activity of the hidden layer for models trained with different objectives.



\ifarxiv
\begin{figure}
\else
\begin{figure}[H]
\fi

\includegraphics[width=0.95\textwidth]{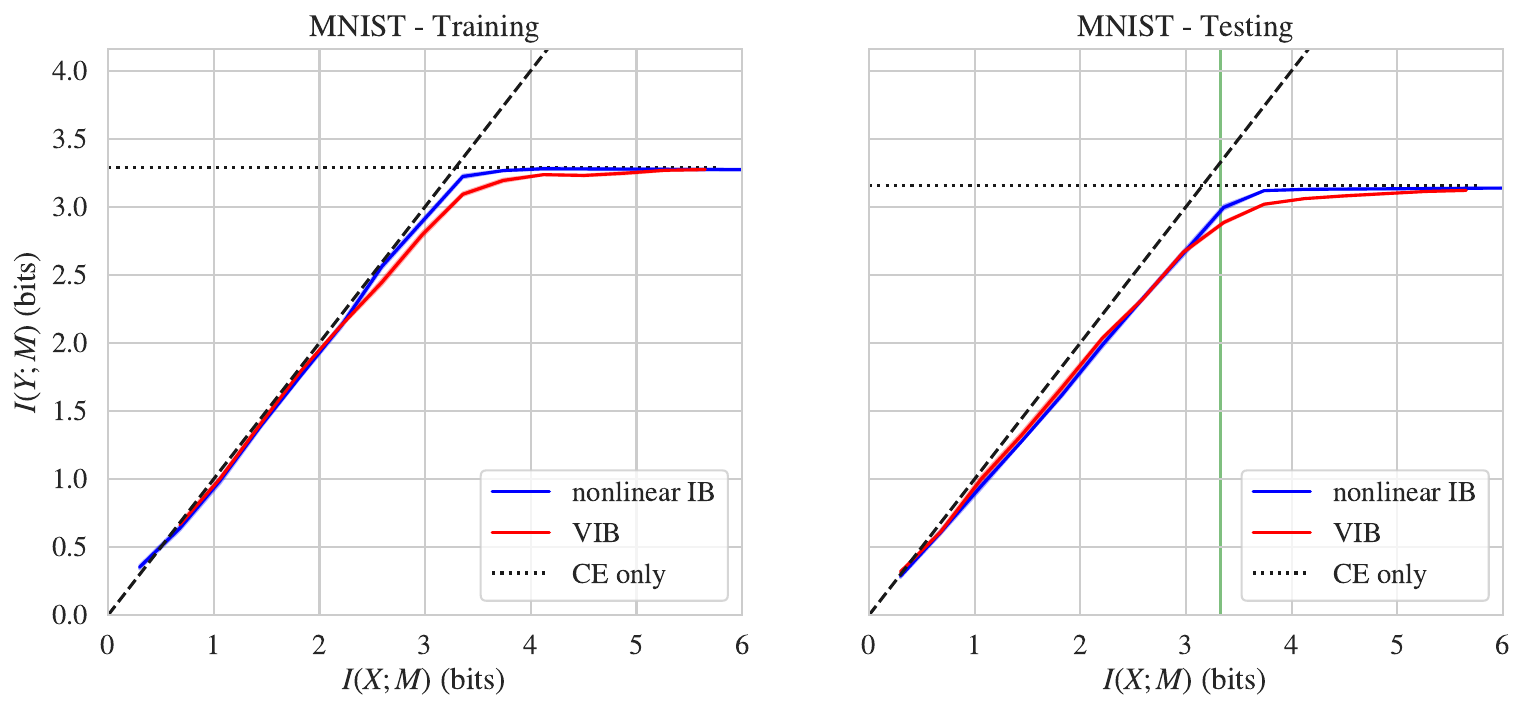}\\
\begin{center}
\includegraphics[width=0.9\textwidth]{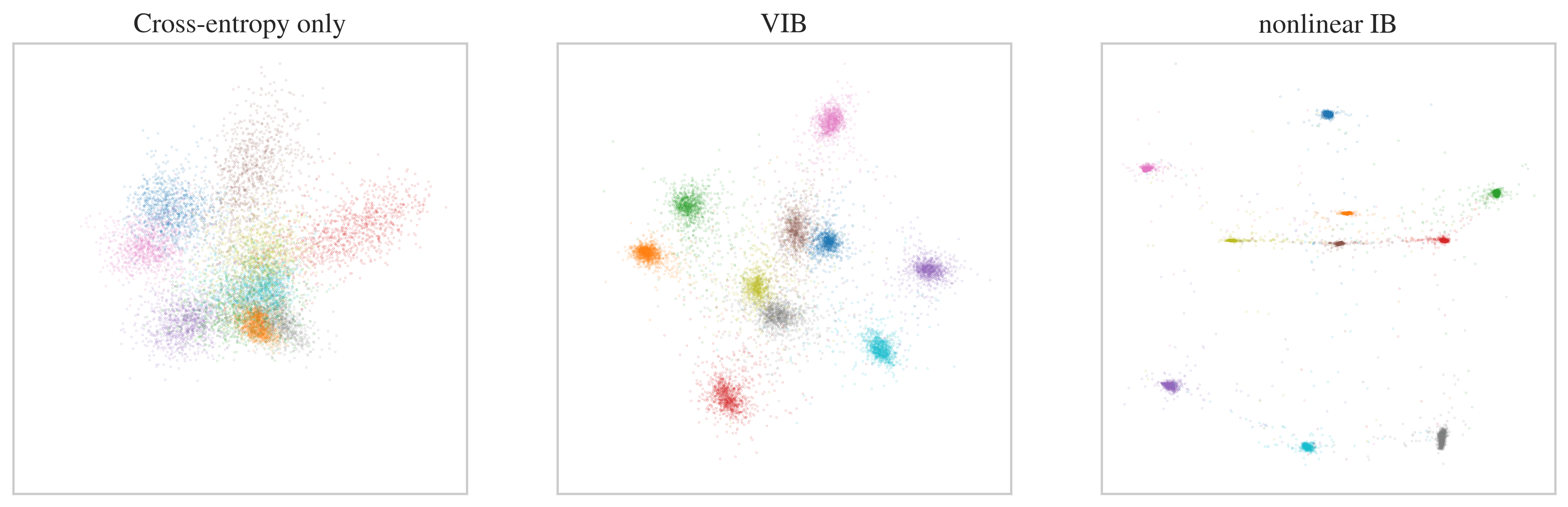}
\end{center}


\caption{{Top row}: Info-plane diagrams for nonlinear IB and variational IB (VIB) on the MNIST training ({\bf left}) and testing ({\bf right}) data. The solid lines indicate means across five runs, 
shaded region indicates the standard error of the mean. The black dashed line is the data-processing inequality bound $\IYM \le \IXM$, the black dotted line indicates the value of $\IYM$ achieved by a baseline model trained only to optimize cross-entropy. 
{Bottom row}: Principal component analysis (PCA) projection of bottleneck layer activity (on testing data, no noise) for models trained with regular cross-entropy loss ({\bf left}), VIB ({\bf middle}), and nonlinear IB ({\bf right}) objectives. The location of the nonlinear IB and VIB models shown in the bottom row is indicated with the green vertical line in the top right panel.\label{fig:mnist-2}
\label{fig:mnist}}

\end{figure}

We first consider the \emph{MNIST} dataset of hand-drawn digits, which contains 60,000 training images and 10,000 testing images.  Each image is 28-by-28 pixels (784 total pixels, so $X \in \mathbb{R}^{784}$), and is classified into 1 of 10 classes corresponding to the digit identity ($Y \in \{1, \dots, 10\}$).  

The top row of Figure \ref{fig:mnist} shows $\IYM$ and $\IXM$ values achieved by  nonlinear IB and VIB on the MNIST dataset.   As can be seen, nonlinear IB achieves better prediction values at the same level of compression than VIB, both on training and testing data.  The difference is especially marked near the ``corner point'' $\IXM = \IYM \approx \log 10$ (which corresponds to maximal compression, given perfect prediction), where nonlinear IB achieved $\approx 0.1$ bits better prediction at the same compression level (see also \cref{tab:results}).   

Further insight is provided by considering the bottleneck representations found when training with nonlinear IB versus VIB versus regular cross-entropy loss. To visualize these bottleneck representations, we selected three models: a baseline model trained only to optimize cross-entropy loss, 
a model trained with nonlinear IB, and 
a model trained with  VIB (the latter two models were chosen to both have $\IXM \approx \log 10$). 
We then measured the activity of their 5-neuron bottleneck hidden layer on the testing dataset, projected down to two dimensions using principal component analysis (PCA).  
Figure \ref{fig:mnist-2} visualizes these two-dimensional projections for these three models, with colors indicating class label (digit identity). 
Training with VIB and nonlinear IB objectives causes inputs corresponding to different digits to fall into well-separated clusters, unlike training with cross-entropy loss. Moreover, the clustering is particularly tight for nonlinear IB, meaning that the bottleneck states carry almost no information about input vectors beyond class identity.  Note that in this regime, where Gaussian components are grouped into tightly separate clusters, our MI upper bound $\MIest$ becomes exact~\cite{kolchinsky_upperbound_2017}.



\ifarxiv
\begin{figure}
\else
\begin{figure}[H]
\fi


\includegraphics[width=0.95\textwidth]{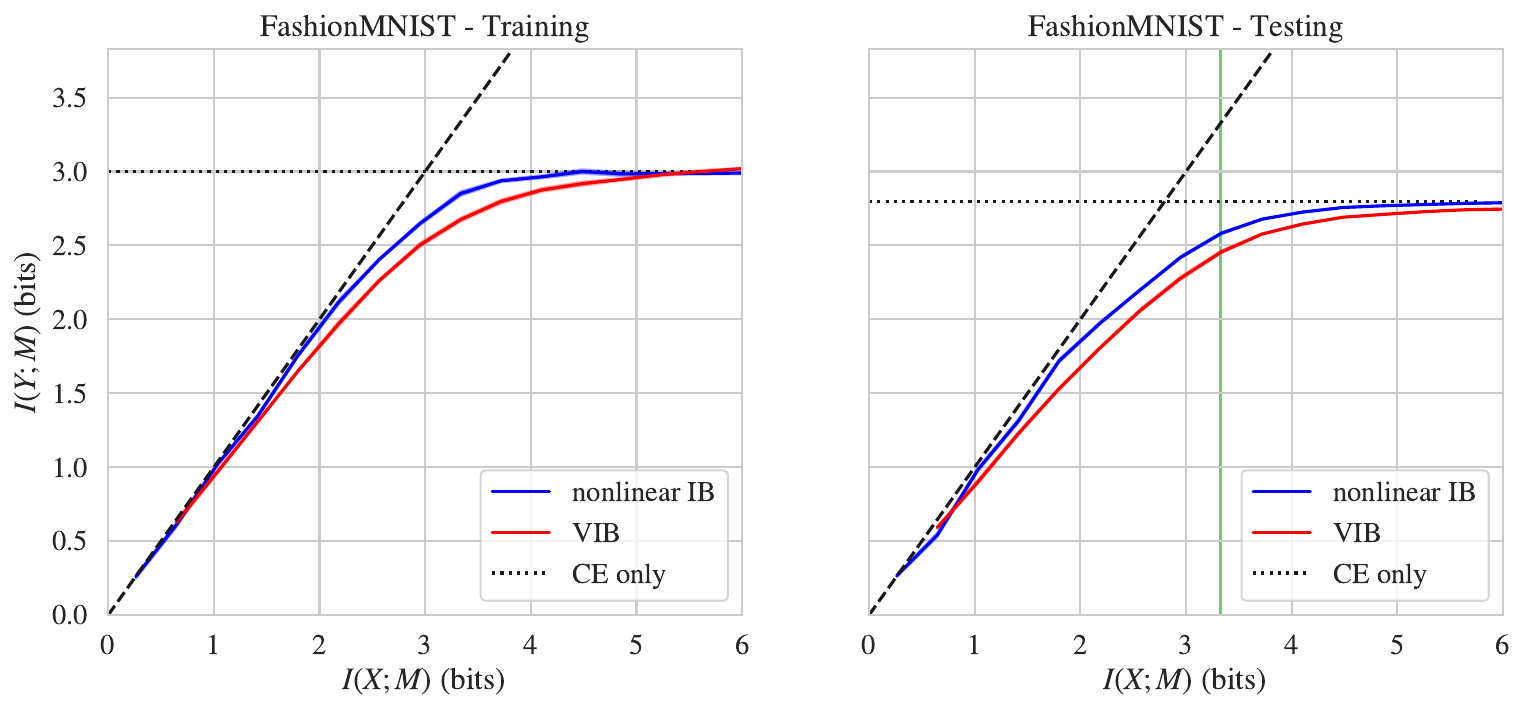}\\
\begin{center}
\includegraphics[width=0.9\textwidth]{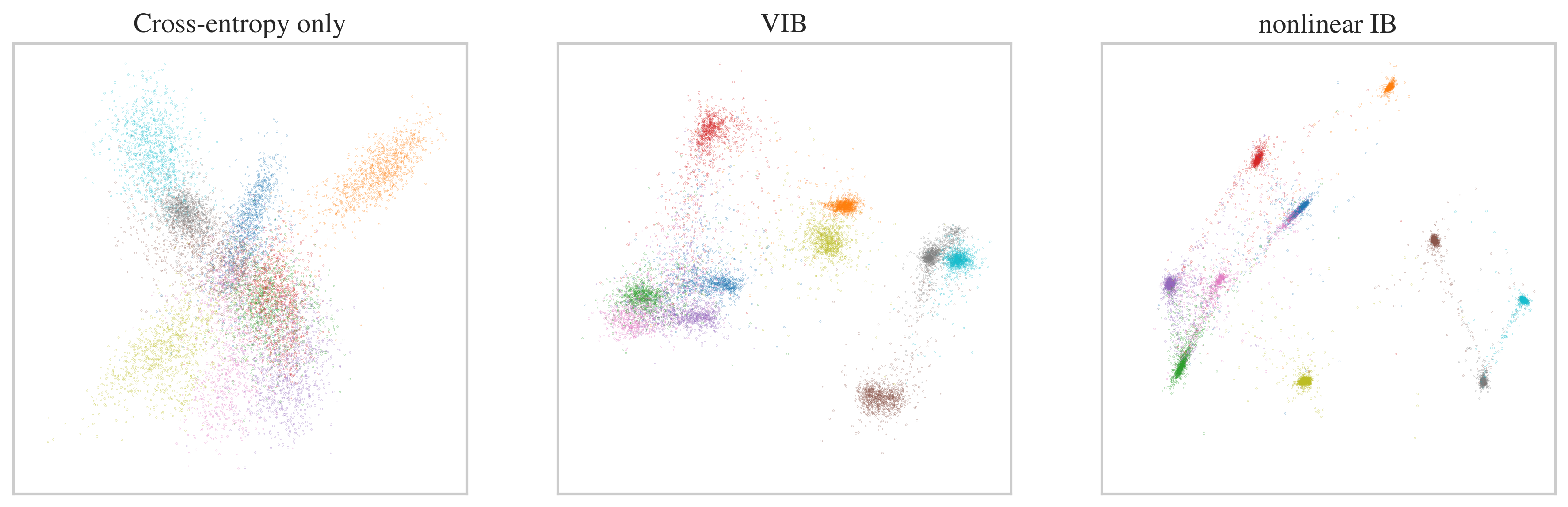}
\end{center}

\caption{Top row: Info-plane  diagrams for nonlinear IB and VIB on the FashionMNIST dataset. 
Bottom row: PCA projection of bottleneck layer activations for models trained only to optimize cross-entropy ({\bf left}), VIB ({\bf middle}), and nonlinear IB ({\bf right}) objectives. 
See caption of Figure \ref{fig:mnist} for details.\label{fig:fashion-mnist}}

\end{figure}

In the next experiment, we considered the recently-proposed \emph{FashionMNIST} dataset. FashionMNIST has the same structure as the MNIST dataset ($28\times 28$ images grouped into 10 classes, with 60,000 training and 10,000 testing images).  Instead of hand-written digits, however, FashionMNIST includes images of clothes  labeled with classes such as ``Dress'', ``Coat'', and ``Sneaker''.  This dataset was designed as a drop-in replacement for MNIST which addresses the problem that MNIST is too easy for modern machine learning (e.g., it is fairly straightforward to achieve $\approx 99\%$ test accuracy on MNIST)~\cite{xiao2017fashion}. FashionMNIST is a more difficult dataset, with typical test accuracies of $\approx 90\%-95\%$.

The top row Figure \ref{fig:fashion-mnist} shows  $\IYM$ and $\IXM$ values achieved by nonlinear IB and VIB on the FashionMNIST dataset. Compared to VIB, nonlinear IB again achieves better prediction values at the same level of compression, both on training and testing data.  The difficulty of FashionMNIST is evident in the fact that neither method gets very close to the corner point $\IXM = \IYM \approx \log 10$.  Nonetheless, nonlinear IB performed better than VIB at a range of compression values, often extracting  $\approx 0.15$  additional bits of prediction at the same compression level (see also \cref{tab:results}).

As for MNIST, we consider the bottleneck representations uncovered when training on FashionMNIST with cross-entropy loss only versus nonlinear IB versus VIB (the latter two models were chosen to have $\IXM \approx \log 10$). 
We measured the activity of the 5-neuron bottleneck  layer on the testing dataset, projected down to two dimensions using PCA.  
The bottom row of {Figure} \ref{fig:fashion-mnist} visualizes these two-dimensional projections for these three models, with colors
 indicating class label (digit identity). 
It can again be seen that models trained with VIB and nonlinear IB map inputs into separated clusters, but that the clusters are significantly tighter for nonlinear IB.

\ifarxiv
\begin{figure*}
\else
\begin{figure*}[t!]
\fi

\includegraphics[width=0.95\textwidth]{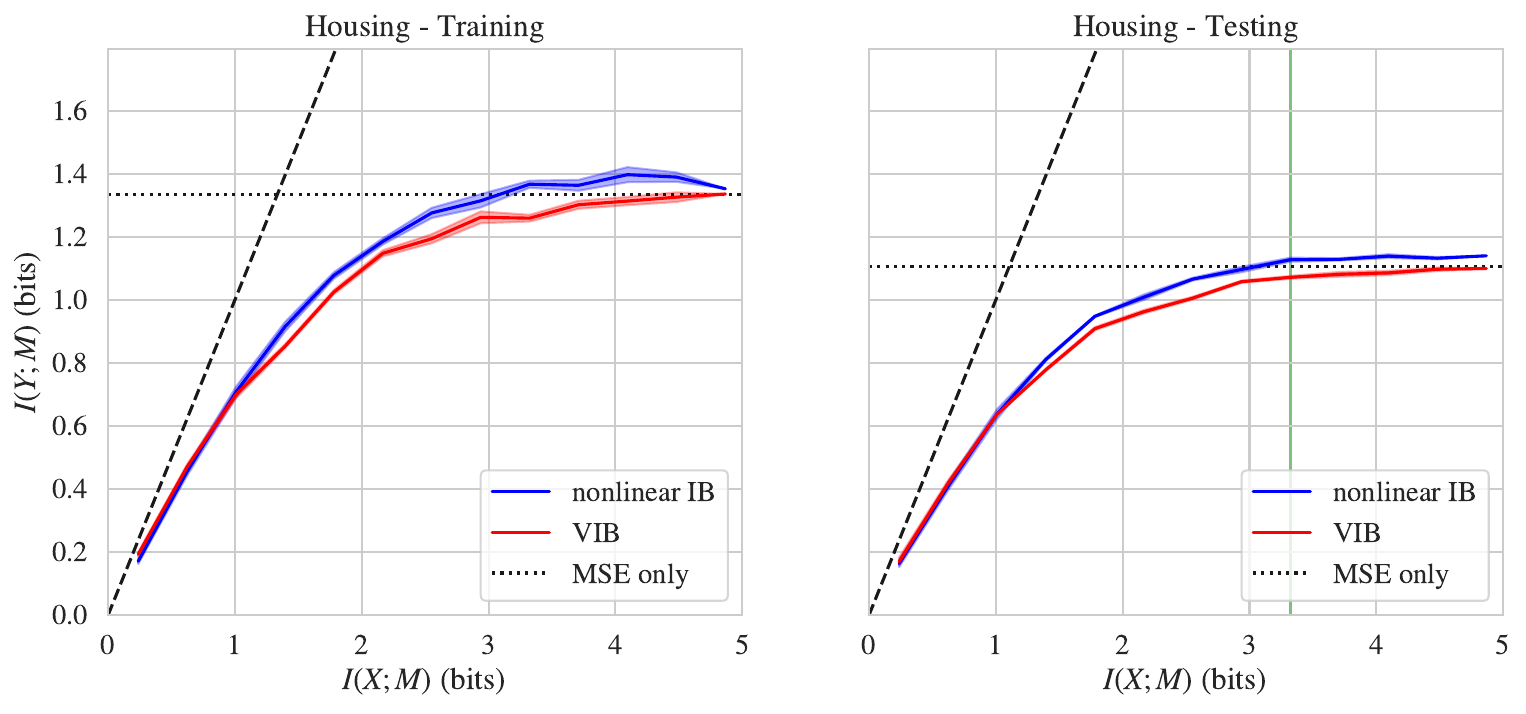}\\
\begin{center}
\includegraphics[width=0.9\textwidth]{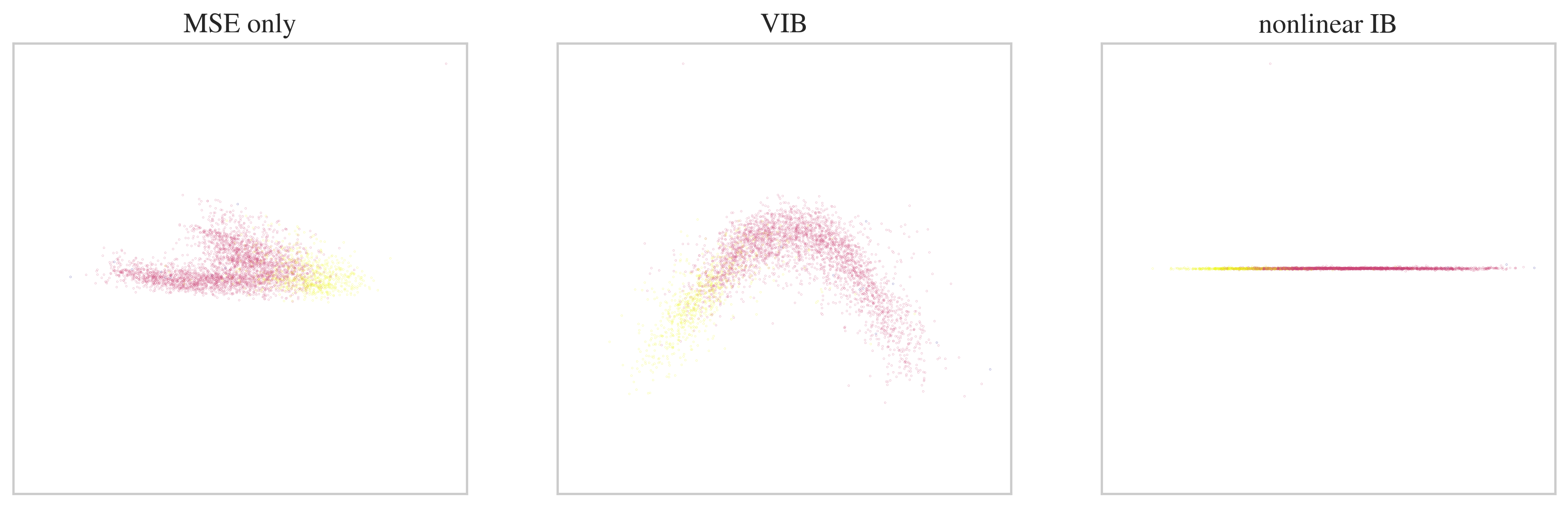}
\end{center}

\caption{Information plane diagrams for nonlinear IB and VIB on the California housing prices dataset. 
Bottom row: PCA projection of bottleneck layer activations for models trained only to optimize mean squared error (MSE) ({\bf left}), VIB ({\bf middle}), and nonlinear IB ({\bf right}) objectives.
See caption of Figure \ref{fig:mnist} for details.\label{fig:housing}}

\end{figure*}

In our final experiment, we considered the {\emph{California housing prices}} 
 dataset.  This is a regression dataset based on the 1990 California census, originally published in~\cite{pace1997sparse} (we use the version distributed with the \texttt{scikit-learn} package~\cite{scikit-learn}). It consists of $N=20,640$ total samples, with one dependent variable (the house price) and 8 independent variables (such as ``longitude'', ``latitude'', and ``number of rooms''). We used the log-transformed house price as the dependent variable $Y$ (this made the distribution of $Y$ closer to a Gaussian). To prepare the training and testing data, we first dropped 992 samples in which the house price was equal to or greater than \$500,000 (prices were clipped at this upper value in the dataset, which distorted the distribution of the dependent variable). We then randomly split the remaining samples into an 80\% training and 20\% testing dataset (the training dataset was then further split into the actual training dataset and a validation dataset, see above).

The top row  of Figure \ref{fig:housing} shows $\IYM$ and $\IXM$ values achieved by  nonlinear IB and VIB on the California housing prices dataset.   Nonlinear IB achieves better prediction values at the same level of compression than VIB, both on training and testing data (see also \cref{tab:results}).  As for the other datasets, we also show the bottleneck representations uncovered when training on California housing prices dataset with MSE loss only versus nonlinear IB versus VIB (the latter two models were chosen to have $\IXM \approx \log 10$). 
The bottom row of Figure \ref{fig:housing} visualizes the two-dimensional PCA projections of bottleneck layer activity for these three models, with colors indicating the dependent variable (log housing price). 
The bottleneck representations uncovered when training with MSE loss only and when training with VIB were somewhat similar. Nonlinear IB, however, finds a different and almost perfectly one-dimensional bottleneck representation. In fact, for the nonlinear IB model, the first principal component explains 99.8\%  of the variance in bottleneck layer activity on testing data. For the models trained with MSE loss and VIB, the first principal component explains  only 76.6\% and 69\% of the variance, respectively. The one-dimensional representation uncovered by nonlinear IB compresses away all information about the input vectors which is not relevant for predicting the dependent variable.

\ifarxiv
\begin{table}
\else
\begin{table}[H]
\fi
\centering
\caption{Amount of prediction $\IYM$ achieved at compression level $\IXM=\log 10$ for both nonlinear IB and VIB.\label{tab:results}}
\begin{tabular}{cccc}
\hline 
\multicolumn{2}{|c|}{\textbf{Dataset}} & \textbf{Nonlinear IB} & \textbf{VIB}\tabularnewline
\hline 
\hline 
\multirow{2}{*}{MNIST} & Training & \textbf{3.22} & 3.09\tabularnewline
\cline{2-4} \cline{3-4} \cline{4-4} 
 & Testing & \textbf{2.99} & 2.88\tabularnewline
\hline 
\multirow{2}{*}{FashionMNIST} & Training & \textbf{2.85} & 2.67\tabularnewline
\cline{2-4} \cline{3-4} \cline{4-4} 
 & Testing & \textbf{2.58} & 2.46\tabularnewline
\hline 
\multirow{2}{*}{California housing} & Training & \textbf{1.37} & 1.26\tabularnewline
\cline{2-4} \cline{3-4} \cline{4-4} 
 & Testing & \textbf{1.13} & 1.07\tabularnewline
\hline 
\end{tabular}

\end{table}

We finish by presenting some of our numerical results in \cref{tab:results}. In particular, we quantify the amount of prediction, $\IYM$, achieved when training with nonlinear IB and VIB at the compression level $\IXM = \log 10$, for training and testing datasets of the three datasets considered above.  Nonlinear IB consistently achieves better prediction at a fixed level of compression. 

\section{Conclusion}
We propose ``nonlinear IB'', a method for exploring the information bottleneck [IB] trade-off curve in a general setting. We allow the input and output variables to be discrete or continuous (though we assume a continuous bottleneck variable). We also allow for  arbitrary (e.g., non-Gaussian) joint distributions over inputs and outputs and for non-linear encoding and decoding maps. We gain this generality by exploiting a new tractable and differentiable bound on the IB objective. 

We describe how to implement our method using off-the-shelf neural network software, and apply it to several standard classification and regression problems. We find that nonlinear IB is able to effectively discover the tradeoff curve, and find solutions that are superior compared with competing methods. We also find that the intermediate representations discovered by nonlinear IB have visibly tighter clusters in the classification problems. In the regression problem, nonlinear IB discovers a one-dimensional intermediate representation. 

We have successfully demonstrated the ability of nonlinear IB to explore the IB curve. It is possible that increased compression may lead to other benefits in supervised learning, such as improved generalization performance or increased robustness to adversarial inputs. 
Exploring its efficacy in these domains remains for future work.

\ifarxiv
\acknowledgments{We thank Steven Van Kuyk and Borja Rodr{\'i}guez G{\'a}lvez for helpful comments. We would also like to thank the Santa Fe Institute for helping to support this research. 
Artemy Kolchinsky and David H. Wolpert were supported by Grant
No. FQXi-RFP-1622 from the FQXi foundation and Grant No. CHE-1648973 from the US National
Science Foundation. Brendan D. Tracey was supported by the AFOSR MURI on multi-information sources of multi-physics systems under Award Number FA9550-15-1-0038.}

\else
\supplementary{ {The following are available online at} \linksupplementary{s1}: Figure S1: Performance of nonlinear IB and VIB when optimizing bounds on regular IB objective.}

\authorcontributions{ {Conceptualization, A.K.;} Funding acquisition, D.H.W.; Software, A.K. and B.D.T.; Visualization, A.K.; Writing-original draft, A.K.; Writing-review \& editing, A.K., B.D.T. and D.H.W. }

\funding{ {This research was} funded by National Science Foundation : CHE-1648973; Foundational Questions Institute : FQXi-RFP-1622; Air Force Office of Scientific Research : A9550-15-1-0038
}

\acknowledgments{We thank Steven Van Kuyk and Borja Rodr{\'i}guez G{\'a}lvez for helpful comments. We would also like to thank the Santa Fe Institute for helping to support this research.
}

\conflictsofinterest{ {The authors declare no conflict of interest.}} 

\fi








\ifarxiv
\bibliography{nonlinearib}
\clearpage
\appendix
\section*{Supplementary Material}

\input{smcontent.tex}

\else
\reftitle{References}



\fi




\end{document}

%% file: smcontent.tex
In the main text, we report the result of optimizing bounds on the ``squared-IB Lagrangian''~\cite{kolchinsky2018caveats}, as defined both in nonlinear IB, 
\begin{align}
\CeLoss - \beta \big[\MIest\big]^2,
\end{align}
and by extending the variational IB (VIB)~\cite{alemi_deep_2016} objective,
\begin{align}
\CeLoss - \beta \big[\DKL(P_\theta(\M|X)\Vert \VM(\M))\big]^2 .
\end{align}
We consider three different datasets (MNIST classification task, FashionMNIST classification task, and the California housing prices regression task). 
In this Supplementary Material, we show results for the same three datasets, but now while optimizing bounds on the regular IB Lagrangian, as defined both by nonlinear IB, 
\begin{align}
\CeLoss - \beta \big[\MIest\big],
\end{align}
and VIB,
\begin{align}
\CeLoss - \beta \big[\DKL(P_\theta(\M|X)\Vert \VM(\M))\big] .
\end{align}
Except for the change in objective, all parameters are the same as those reported in the main text.  See the caption of Figure 1 in the main text for details on how to interpret the following info-plane diagrams.

\clearpage

\begin{center}
\includegraphics[width=0.93\textwidth]{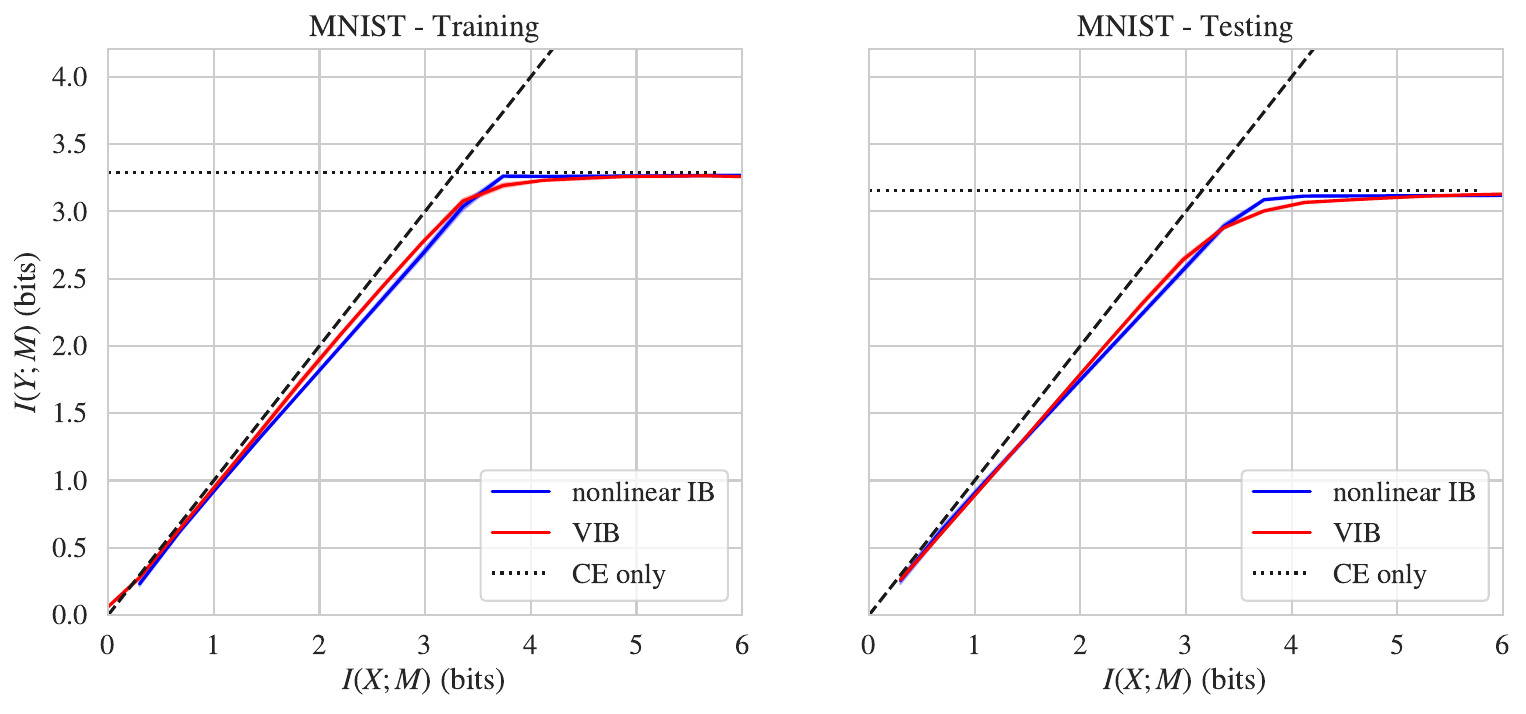}

\includegraphics[width=0.93\textwidth]{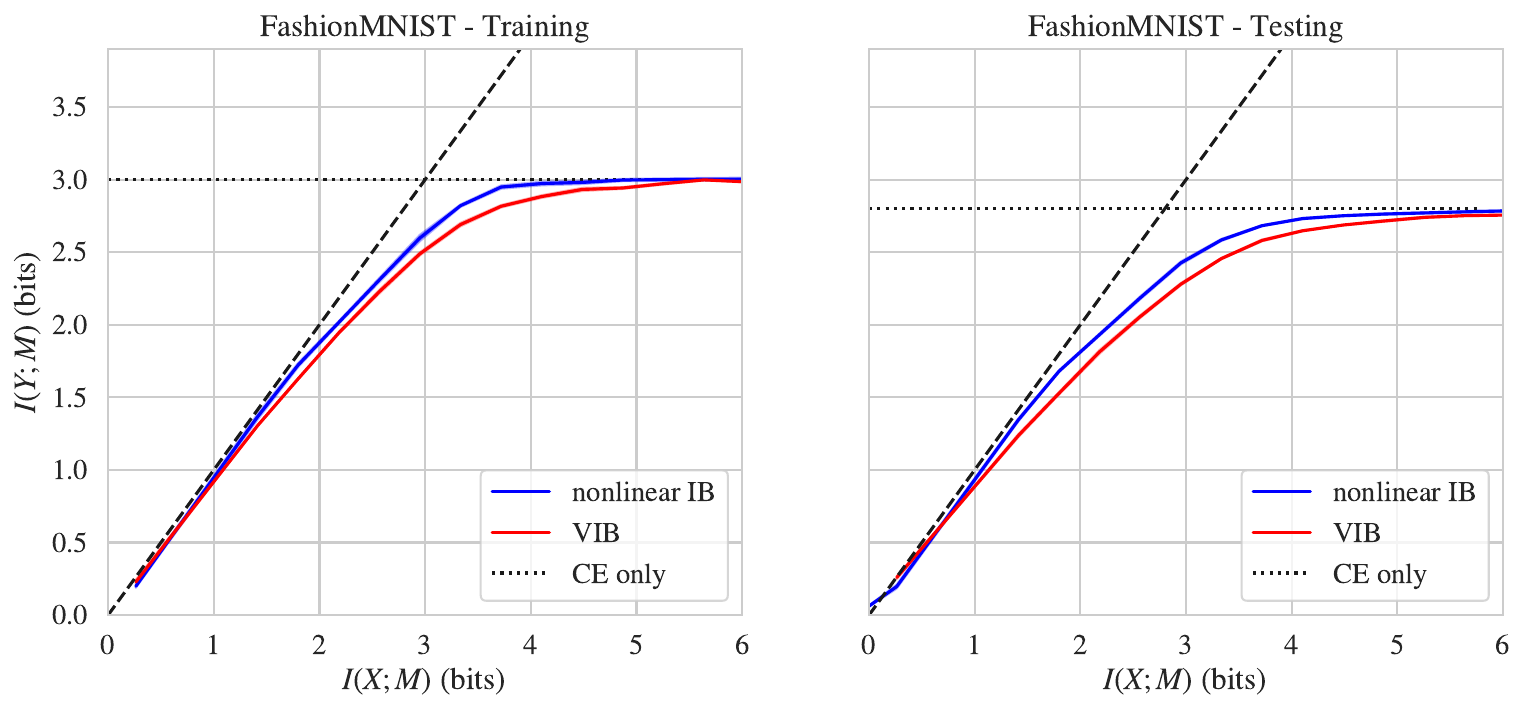}

\includegraphics[width=0.93\textwidth]{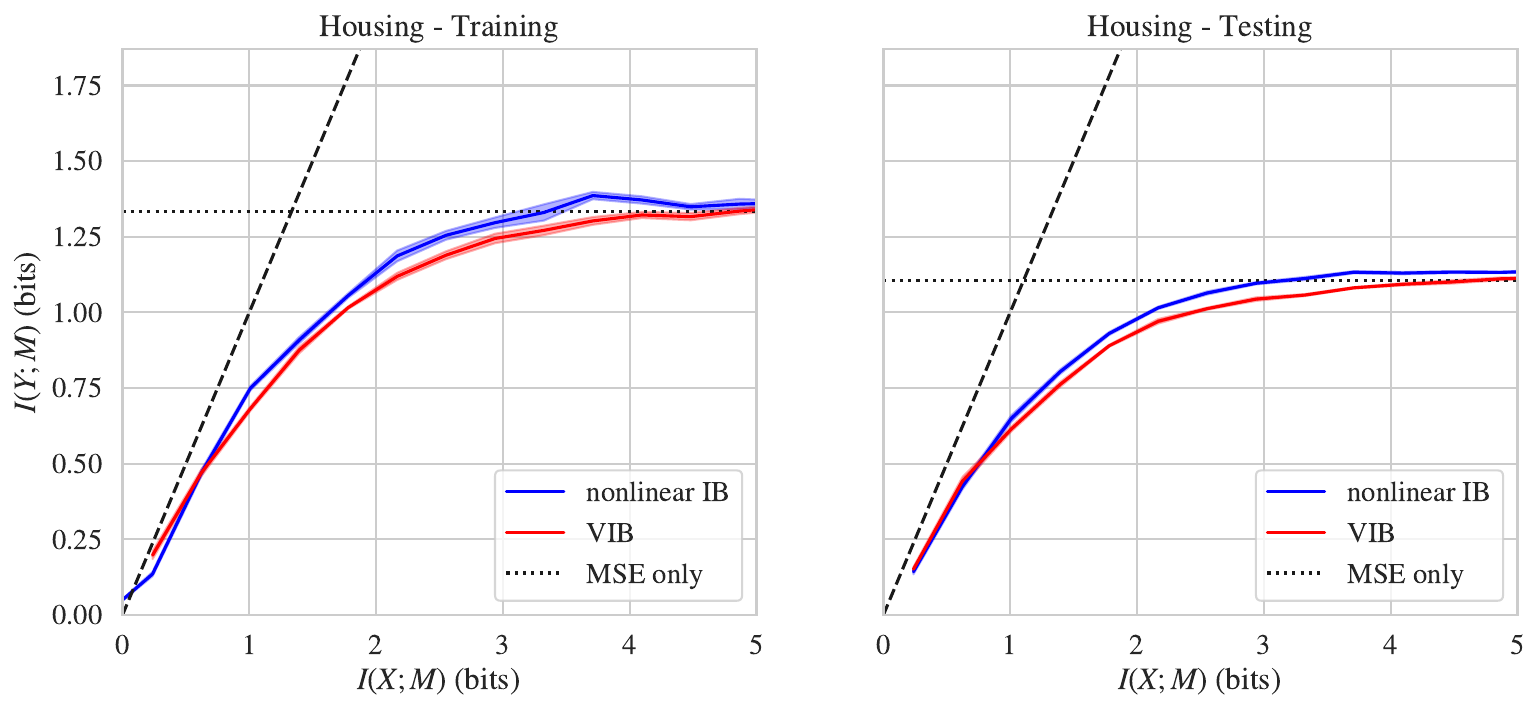}

Figure S1: Performance of nonlinear IB and VIB when optimizing bounds on regular IB objective
\end{center}

%% file: main.bbl
\begin{thebibliography}{57}%
\makeatletter
\providecommand \@ifxundefined [1]{%
 \@ifx{#1\undefined}
}%
\providecommand \@ifnum [1]{%
 \ifnum #1\expandafter \@firstoftwo
 \else \expandafter \@secondoftwo
 \fi
}%
\providecommand \@ifx [1]{%
 \ifx #1\expandafter \@firstoftwo
 \else \expandafter \@secondoftwo
 \fi
}%
\providecommand \natexlab [1]{#1}%
\providecommand \enquote  [1]{``#1''}%
\providecommand \bibnamefont  [1]{#1}%
\providecommand \bibfnamefont [1]{#1}%
\providecommand \citenamefont [1]{#1}%
\providecommand \href@noop [0]{\@secondoftwo}%
\providecommand \href [0]{\begingroup \@sanitize@url \@href}%
\providecommand \@href[1]{\@@startlink{#1}\@@href}%
\providecommand \@@href[1]{\endgroup#1\@@endlink}%
\providecommand \@sanitize@url [0]{\catcode `\\12\catcode `\$12\catcode
  `\&12\catcode `\#12\catcode `\^12\catcode `\_12\catcode `\%12\relax}%
\providecommand \@@startlink[1]{}%
\providecommand \@@endlink[0]{}%
\providecommand \url  [0]{\begingroup\@sanitize@url \@url }%
\providecommand \@url [1]{\endgroup\@href {#1}{\urlprefix }}%
\providecommand \urlprefix  [0]{URL }%
\providecommand \Eprint [0]{\href }%
\providecommand \doibase [0]{http://dx.doi.org/}%
\providecommand \selectlanguage [0]{\@gobble}%
\providecommand \bibinfo  [0]{\@secondoftwo}%
\providecommand \bibfield  [0]{\@secondoftwo}%
\providecommand \translation [1]{[#1]}%
\providecommand \BibitemOpen [0]{}%
\providecommand \bibitemStop [0]{}%
\providecommand \bibitemNoStop [0]{.\EOS\space}%
\providecommand \EOS [0]{\spacefactor3000\relax}%
\providecommand \BibitemShut  [1]{\csname bibitem#1\endcsname}%
\let\auto@bib@innerbib\@empty
\bibitem [{\citenamefont {Tishby}\ \emph {et~al.}(1999)\citenamefont {Tishby},
  \citenamefont {Pereira},\ and\ \citenamefont
  {Bialek}}]{tishby_information_1999}%
  \BibitemOpen
  \bibfield  {author} {\bibinfo {author} {\bibfnamefont {N.}~\bibnamefont
  {Tishby}}, \bibinfo {author} {\bibfnamefont {F.}~\bibnamefont {Pereira}}, \
  and\ \bibinfo {author} {\bibfnamefont {W.}~\bibnamefont {Bialek}},\
  }\bibfield  {title} {\enquote {\bibinfo {title} {The information bottleneck
  method},}\ }in\ \href@noop {} {\emph {\bibinfo {booktitle} {37th {Allerton}
  {Conf} on {Communication}}}}\ (\bibinfo {year} {1999})\BibitemShut {NoStop}%
\bibitem [{\citenamefont {Dimitrov}\ and\ \citenamefont
  {Miller}(2001)}]{dimitrov_neural_2001}%
  \BibitemOpen
  \bibfield  {author} {\bibinfo {author} {\bibfnamefont {Alexander~G.}\
  \bibnamefont {Dimitrov}}\ and\ \bibinfo {author} {\bibfnamefont {John~P.}\
  \bibnamefont {Miller}},\ }\bibfield  {title} {\enquote {\bibinfo {title}
  {Neural coding and decoding: communication channels and quantization},}\
  }\href@noop {} {\bibfield  {journal} {\bibinfo  {journal} {Network:
  Computation in Neural Systems}\ }\textbf {\bibinfo {volume} {12}},\ \bibinfo
  {pages} {441--472} (\bibinfo {year} {2001})}\BibitemShut {NoStop}%
\bibitem [{\citenamefont {Samengo}(2002)}]{samengo_information_2002}%
  \BibitemOpen
  \bibfield  {author} {\bibinfo {author} {\bibfnamefont {Inés}\ \bibnamefont
  {Samengo}},\ }\bibfield  {title} {\enquote {\bibinfo {title} {Information
  loss in an optimal maximum likelihood decoding},}\ }\href@noop {} {\bibfield
  {journal} {\bibinfo  {journal} {Neural computation}\ }\textbf {\bibinfo
  {volume} {14}},\ \bibinfo {pages} {771--779} (\bibinfo {year}
  {2002})}\BibitemShut {NoStop}%
\bibitem [{\citenamefont {Witsenhausen}\ and\ \citenamefont
  {Wyner}(1975)}]{witsenhausen_conditional_1975}%
  \BibitemOpen
  \bibfield  {author} {\bibinfo {author} {\bibfnamefont {H.}~\bibnamefont
  {Witsenhausen}}\ and\ \bibinfo {author} {\bibfnamefont {A.}~\bibnamefont
  {Wyner}},\ }\bibfield  {title} {\enquote {\bibinfo {title} {A conditional
  entropy bound for a pair of discrete random variables},}\ }\href@noop {}
  {\bibfield  {journal} {\bibinfo  {journal} {IEEE Transactions on Information
  Theory}\ }\textbf {\bibinfo {volume} {21}},\ \bibinfo {pages} {493--501}
  (\bibinfo {year} {1975})}\BibitemShut {NoStop}%
\bibitem [{\citenamefont {Ahlswede}\ and\ \citenamefont
  {Körner}(1975)}]{ahlswede_source_1975}%
  \BibitemOpen
  \bibfield  {author} {\bibinfo {author} {\bibfnamefont {Rudolf}\ \bibnamefont
  {Ahlswede}}\ and\ \bibinfo {author} {\bibfnamefont {János}\ \bibnamefont
  {Körner}},\ }\bibfield  {title} {\enquote {\bibinfo {title} {Source coding
  with {Side} {Information} and a {Converse} for {Degraded} {Broadcast
  Channels}},}\ }\href@noop {} {\bibfield  {journal} {\bibinfo  {journal} {IEEE
  Transaction on Information Theory}\ ,\ \bibinfo {pages} {9}} (\bibinfo {year}
  {1975})}\BibitemShut {NoStop}%
\bibitem [{\citenamefont {Gilad-Bachrach}\ \emph {et~al.}(2003)\citenamefont
  {Gilad-Bachrach}, \citenamefont {Navot},\ and\ \citenamefont
  {Tishby}}]{goos_information_2003}%
  \BibitemOpen
  \bibfield  {author} {\bibinfo {author} {\bibfnamefont {Ran}\ \bibnamefont
  {Gilad-Bachrach}}, \bibinfo {author} {\bibfnamefont {Amir}\ \bibnamefont
  {Navot}}, \ and\ \bibinfo {author} {\bibfnamefont {Naftali}\ \bibnamefont
  {Tishby}},\ }\bibfield  {title} {\enquote {\bibinfo {title} {An {Information}
  {Theoretic} {Tradeoff} between {Complexity} and {Accuracy}},}\ }in\
  \href@noop {} {\emph {\bibinfo {booktitle} {Learning {Theory} and {Kernel}
  {Machines}}}},\ Vol.\ \bibinfo {volume} {2777},\ \bibinfo {editor} {edited
  by\ \bibinfo {editor} {\bibfnamefont {Gerhard}\ \bibnamefont {Goos}},
  \bibinfo {editor} {\bibfnamefont {Juris}\ \bibnamefont {Hartmanis}}, \bibinfo
  {editor} {\bibfnamefont {Jan}\ \bibnamefont {van Leeuwen}}, \bibinfo {editor}
  {\bibfnamefont {Bernhard}\ \bibnamefont {Schölkopf}}, \ and\ \bibinfo
  {editor} {\bibfnamefont {Manfred~K.}\ \bibnamefont {Warmuth}}}\ (\bibinfo
  {publisher} {Springer Berlin Heidelberg},\ \bibinfo {year} {2003})\ pp.\
  \bibinfo {pages} {595--609}\BibitemShut {NoStop}%
\bibitem [{\citenamefont {Slonim}\ and\ \citenamefont
  {Tishby}(2000)}]{slonim2000document}%
  \BibitemOpen
  \bibfield  {author} {\bibinfo {author} {\bibfnamefont {Noam}\ \bibnamefont
  {Slonim}}\ and\ \bibinfo {author} {\bibfnamefont {Naftali}\ \bibnamefont
  {Tishby}},\ }\bibfield  {title} {\enquote {\bibinfo {title} {Document
  clustering using word clusters via the information bottleneck method},}\ }in\
  \href@noop {} {\emph {\bibinfo {booktitle} {Proceedings of the 23rd annual
  international ACM SIGIR conference on Research and development in information
  retrieval}}}\ (\bibinfo {organization} {ACM},\ \bibinfo {year} {2000})\ pp.\
  \bibinfo {pages} {208--215}\BibitemShut {NoStop}%
\bibitem [{\citenamefont {Tishby}\ and\ \citenamefont
  {Slonim}(2001)}]{tishby_data_2001}%
  \BibitemOpen
  \bibfield  {author} {\bibinfo {author} {\bibfnamefont {Naftali}\ \bibnamefont
  {Tishby}}\ and\ \bibinfo {author} {\bibfnamefont {Noam}\ \bibnamefont
  {Slonim}},\ }\bibfield  {title} {\enquote {\bibinfo {title} {Data clustering
  by markovian relaxation and the information bottleneck method},}\ }in\
  \href@noop {} {\emph {\bibinfo {booktitle} {Advances in neural information
  processing systems}}}\ (\bibinfo {year} {2001})\ pp.\ \bibinfo {pages}
  {640--646}\BibitemShut {NoStop}%
\bibitem [{\citenamefont {Cardinal}(2003)}]{cardinal2003compression}%
  \BibitemOpen
  \bibfield  {author} {\bibinfo {author} {\bibfnamefont {Jean}\ \bibnamefont
  {Cardinal}},\ }\bibfield  {title} {\enquote {\bibinfo {title} {Compression of
  side information},}\ }in\ \href@noop {} {\emph {\bibinfo {booktitle}
  {icme}}}\ (\bibinfo {organization} {IEEE},\ \bibinfo {year} {2003})\ pp.\
  \bibinfo {pages} {569--572}\BibitemShut {NoStop}%
\bibitem [{\citenamefont {Zeitler}\ \emph {et~al.}(2008)\citenamefont
  {Zeitler}, \citenamefont {Koetter}, \citenamefont {Bauch},\ and\
  \citenamefont {Widmer}}]{zeitler2008design}%
  \BibitemOpen
  \bibfield  {author} {\bibinfo {author} {\bibfnamefont {Georg}\ \bibnamefont
  {Zeitler}}, \bibinfo {author} {\bibfnamefont {Ralf}\ \bibnamefont {Koetter}},
  \bibinfo {author} {\bibfnamefont {Gerhard}\ \bibnamefont {Bauch}}, \ and\
  \bibinfo {author} {\bibfnamefont {Joerg}\ \bibnamefont {Widmer}},\ }\bibfield
   {title} {\enquote {\bibinfo {title} {Design of network coding functions in
  multihop relay networks},}\ }in\ \href@noop {} {\emph {\bibinfo {booktitle}
  {Turbo Codes and Related Topics, 2008 5th International Symposium on}}}\
  (\bibinfo {organization} {Citeseer},\ \bibinfo {year} {2008})\ pp.\ \bibinfo
  {pages} {249--254}\BibitemShut {NoStop}%
\bibitem [{\citenamefont {Courtade}\ and\ \citenamefont
  {Wesel}(2011)}]{courtade2011multiterminal}%
  \BibitemOpen
  \bibfield  {author} {\bibinfo {author} {\bibfnamefont {Thomas~A}\
  \bibnamefont {Courtade}}\ and\ \bibinfo {author} {\bibfnamefont {Richard~D}\
  \bibnamefont {Wesel}},\ }\bibfield  {title} {\enquote {\bibinfo {title}
  {Multiterminal source coding with an entropy-based distortion measure},}\
  }in\ \href@noop {} {\emph {\bibinfo {booktitle} {Information Theory
  Proceedings (ISIT), 2011 IEEE International Symposium on}}}\ (\bibinfo
  {organization} {IEEE},\ \bibinfo {year} {2011})\ pp.\ \bibinfo {pages}
  {2040--2044}\BibitemShut {NoStop}%
\bibitem [{\citenamefont {Lazebnik}\ and\ \citenamefont
  {Raginsky}(2008)}]{lazebnik2008supervised}%
  \BibitemOpen
  \bibfield  {author} {\bibinfo {author} {\bibfnamefont {Svetlana}\
  \bibnamefont {Lazebnik}}\ and\ \bibinfo {author} {\bibfnamefont {Maxim}\
  \bibnamefont {Raginsky}},\ }\bibfield  {title} {\enquote {\bibinfo {title}
  {Supervised learning of quantizer codebooks by information loss
  minimization},}\ }\href@noop {} {\bibfield  {journal} {\bibinfo  {journal}
  {IEEE transactions on pattern analysis and machine intelligence}\ }\textbf
  {\bibinfo {volume} {31}},\ \bibinfo {pages} {1294--1309} (\bibinfo {year}
  {2008})}\BibitemShut {NoStop}%
\bibitem [{\citenamefont {Winn}\ \emph {et~al.}(2005)\citenamefont {Winn},
  \citenamefont {Criminisi},\ and\ \citenamefont {Minka}}]{winn2005object}%
  \BibitemOpen
  \bibfield  {author} {\bibinfo {author} {\bibfnamefont {J}~\bibnamefont
  {Winn}}, \bibinfo {author} {\bibfnamefont {A}~\bibnamefont {Criminisi}}, \
  and\ \bibinfo {author} {\bibfnamefont {T}~\bibnamefont {Minka}},\ }\bibfield
  {title} {\enquote {\bibinfo {title} {Object categorization by learned
  universal visual dictionary},}\ }in\ \href@noop {} {\emph {\bibinfo
  {booktitle} {Tenth IEEE International Conference on Computer Vision (ICCV'05)
  Volume 1}}},\ Vol.~\bibinfo {volume} {2}\ (\bibinfo {organization} {IEEE},\
  \bibinfo {year} {2005})\ pp.\ \bibinfo {pages} {1800--1807}\BibitemShut
  {NoStop}%
\bibitem [{\citenamefont {Hecht}\ \emph {et~al.}(2009)\citenamefont {Hecht},
  \citenamefont {Noor},\ and\ \citenamefont {Tishby}}]{hecht2009speaker}%
  \BibitemOpen
  \bibfield  {author} {\bibinfo {author} {\bibfnamefont {Ron~M}\ \bibnamefont
  {Hecht}}, \bibinfo {author} {\bibfnamefont {Elad}\ \bibnamefont {Noor}}, \
  and\ \bibinfo {author} {\bibfnamefont {Naftali}\ \bibnamefont {Tishby}},\
  }\bibfield  {title} {\enquote {\bibinfo {title} {Speaker recognition by
  gaussian information bottleneck},}\ }in\ \href@noop {} {\emph {\bibinfo
  {booktitle} {Tenth Annual Conference of the International Speech
  Communication Association}}}\ (\bibinfo {year} {2009})\BibitemShut {NoStop}%
\bibitem [{\citenamefont {Yaman}\ \emph {et~al.}(2012)\citenamefont {Yaman},
  \citenamefont {Pelecanos},\ and\ \citenamefont
  {Sarikaya}}]{yaman2012bottleneck}%
  \BibitemOpen
  \bibfield  {author} {\bibinfo {author} {\bibfnamefont {Sibel}\ \bibnamefont
  {Yaman}}, \bibinfo {author} {\bibfnamefont {Jason}\ \bibnamefont
  {Pelecanos}}, \ and\ \bibinfo {author} {\bibfnamefont {Ruhi}\ \bibnamefont
  {Sarikaya}},\ }\bibfield  {title} {\enquote {\bibinfo {title} {Bottleneck
  features for speaker recognition},}\ }in\ \href@noop {} {\emph {\bibinfo
  {booktitle} {Odyssey 2012-The Speaker and Language Recognition Workshop}}}\
  (\bibinfo {year} {2012})\BibitemShut {NoStop}%
\bibitem [{\citenamefont {Van~Kuyk}\ \emph {et~al.}(2017)\citenamefont
  {Van~Kuyk}, \citenamefont {Kleijn},\ and\ \citenamefont
  {Hendriks}}]{van2017information}%
  \BibitemOpen
  \bibfield  {author} {\bibinfo {author} {\bibfnamefont {Steven}\ \bibnamefont
  {Van~Kuyk}}, \bibinfo {author} {\bibfnamefont {W~Bastiaan}\ \bibnamefont
  {Kleijn}}, \ and\ \bibinfo {author} {\bibfnamefont {Richard~C}\ \bibnamefont
  {Hendriks}},\ }\bibfield  {title} {\enquote {\bibinfo {title} {On the
  information rate of speech communication},}\ }in\ \href@noop {} {\emph
  {\bibinfo {booktitle} {2017 IEEE International Conference on Acoustics,
  Speech and Signal Processing (ICASSP)}}}\ (\bibinfo {organization} {IEEE},\
  \bibinfo {year} {2017})\ pp.\ \bibinfo {pages} {5625--5629}\BibitemShut
  {NoStop}%
\bibitem [{\citenamefont {Van~Kuyk}(2019)}]{van2019speech}%
  \BibitemOpen
  \bibfield  {author} {\bibinfo {author} {\bibfnamefont {Steven}\ \bibnamefont
  {Van~Kuyk}},\ }\emph {\bibinfo {title} {Speech Communication from an
  Information Theoretical Perspective}},\ \href@noop {} {Ph.D. thesis}
  (\bibinfo {year} {2019})\BibitemShut {NoStop}%
\bibitem [{\citenamefont {Zaslavsky}\ \emph {et~al.}(2018)\citenamefont
  {Zaslavsky}, \citenamefont {Kemp}, \citenamefont {Regier},\ and\
  \citenamefont {Tishby}}]{zaslavsky2018efficient}%
  \BibitemOpen
  \bibfield  {author} {\bibinfo {author} {\bibfnamefont {Noga}\ \bibnamefont
  {Zaslavsky}}, \bibinfo {author} {\bibfnamefont {Charles}\ \bibnamefont
  {Kemp}}, \bibinfo {author} {\bibfnamefont {Terry}\ \bibnamefont {Regier}}, \
  and\ \bibinfo {author} {\bibfnamefont {Naftali}\ \bibnamefont {Tishby}},\
  }\bibfield  {title} {\enquote {\bibinfo {title} {Efficient compression in
  color naming and its evolution},}\ }\href@noop {} {\bibfield  {journal}
  {\bibinfo  {journal} {Proceedings of the National Academy of Sciences}\
  }\textbf {\bibinfo {volume} {115}},\ \bibinfo {pages} {7937--7942} (\bibinfo
  {year} {2018})}\BibitemShut {NoStop}%
\bibitem [{\citenamefont
  {Rodr{\'i}guez~G{\'a}lvez}(2019)}]{rodriguez_galvez_information_2019}%
  \BibitemOpen
  \bibfield  {author} {\bibinfo {author} {\bibfnamefont {Borja}\ \bibnamefont
  {Rodr{\'i}guez~G{\'a}lvez}},\ }\href
  {http://urn.kb.se/resolve?urn=urn:nbn:se:kth:diva-254421} {\emph {\bibinfo
  {title} {The {Information} {Bottleneck} : {Connections} to {Other}
  {Problems}, {Learning} and {Exploration} of the {IB} {Curve}}}}\ (\bibinfo
  {year} {2019})\BibitemShut {NoStop}%
\bibitem [{\citenamefont {Hafez-Kolahi}\ and\ \citenamefont
  {Kasaei}(2019)}]{hafez2019information}%
  \BibitemOpen
  \bibfield  {author} {\bibinfo {author} {\bibfnamefont {Hassan}\ \bibnamefont
  {Hafez-Kolahi}}\ and\ \bibinfo {author} {\bibfnamefont {Shohreh}\
  \bibnamefont {Kasaei}},\ }\bibfield  {title} {\enquote {\bibinfo {title}
  {Information bottleneck and its applications in deep learning},}\ }\href@noop
  {} {\bibfield  {journal} {\bibinfo  {journal} {arXiv preprint
  arXiv:1904.03743}\ } (\bibinfo {year} {2019})}\BibitemShut {NoStop}%
\bibitem [{\citenamefont {Tishby}\ and\ \citenamefont
  {Zaslavsky}(2015)}]{tishby2015deep}%
  \BibitemOpen
  \bibfield  {author} {\bibinfo {author} {\bibfnamefont {Naftali}\ \bibnamefont
  {Tishby}}\ and\ \bibinfo {author} {\bibfnamefont {Noga}\ \bibnamefont
  {Zaslavsky}},\ }\bibfield  {title} {\enquote {\bibinfo {title} {Deep learning
  and the information bottleneck principle},}\ }in\ \href@noop {} {\emph
  {\bibinfo {booktitle} {Information Theory Workshop (ITW), 2015 IEEE}}}\
  (\bibinfo {organization} {IEEE},\ \bibinfo {year} {2015})\ pp.\ \bibinfo
  {pages} {1--5}\BibitemShut {NoStop}%
\bibitem [{\citenamefont {Shamir}\ \emph {et~al.}(2010)\citenamefont {Shamir},
  \citenamefont {Sabato},\ and\ \citenamefont {Tishby}}]{shamir2010learning}%
  \BibitemOpen
  \bibfield  {author} {\bibinfo {author} {\bibfnamefont {Ohad}\ \bibnamefont
  {Shamir}}, \bibinfo {author} {\bibfnamefont {Sivan}\ \bibnamefont {Sabato}},
  \ and\ \bibinfo {author} {\bibfnamefont {Naftali}\ \bibnamefont {Tishby}},\
  }\bibfield  {title} {\enquote {\bibinfo {title} {Learning and generalization
  with the information bottleneck},}\ }\href@noop {} {\bibfield  {journal}
  {\bibinfo  {journal} {Theoretical Computer Science}\ }\textbf {\bibinfo
  {volume} {411}},\ \bibinfo {pages} {2696--2711} (\bibinfo {year}
  {2010})}\BibitemShut {NoStop}%
\bibitem [{\citenamefont {Vera}\ \emph {et~al.}(2018)\citenamefont {Vera},
  \citenamefont {Piantanida},\ and\ \citenamefont {Vega}}]{vera2018role}%
  \BibitemOpen
  \bibfield  {author} {\bibinfo {author} {\bibfnamefont {Matias}\ \bibnamefont
  {Vera}}, \bibinfo {author} {\bibfnamefont {Pablo}\ \bibnamefont
  {Piantanida}}, \ and\ \bibinfo {author} {\bibfnamefont {Leonardo~Rey}\
  \bibnamefont {Vega}},\ }\bibfield  {title} {\enquote {\bibinfo {title} {The
  role of the information bottleneck in representation learning},}\ }in\
  \href@noop {} {\emph {\bibinfo {booktitle} {2018 IEEE International Symposium
  on Information Theory (ISIT)}}}\ (\bibinfo {organization} {IEEE},\ \bibinfo
  {year} {2018})\ pp.\ \bibinfo {pages} {1580--1584}\BibitemShut {NoStop}%
\bibitem [{\citenamefont {Alemi}\ \emph {et~al.}(2017)\citenamefont {Alemi},
  \citenamefont {Fischer}, \citenamefont {Dillon},\ and\ \citenamefont
  {Murphy}}]{alemi_deep_2016}%
  \BibitemOpen
  \bibfield  {author} {\bibinfo {author} {\bibfnamefont {Alexander~A.}\
  \bibnamefont {Alemi}}, \bibinfo {author} {\bibfnamefont {Ian}\ \bibnamefont
  {Fischer}}, \bibinfo {author} {\bibfnamefont {Joshua~V.}\ \bibnamefont
  {Dillon}}, \ and\ \bibinfo {author} {\bibfnamefont {Kevin}\ \bibnamefont
  {Murphy}},\ }\bibfield  {title} {\enquote {\bibinfo {title} {Deep
  {Variational} {Information} {Bottleneck}},}\ }in\ \href@noop {} {\emph
  {\bibinfo {booktitle} {International Conference on Learning Representations
  (ICLR)}}}\ (\bibinfo {year} {2017})\BibitemShut {NoStop}%
\bibitem [{\citenamefont {Alemi}\ \emph
  {et~al.}(2018{\natexlab{a}})\citenamefont {Alemi}, \citenamefont {Fischer},\
  and\ \citenamefont {Dillon}}]{alemi2018uncertainty}%
  \BibitemOpen
  \bibfield  {author} {\bibinfo {author} {\bibfnamefont {Alexander~A}\
  \bibnamefont {Alemi}}, \bibinfo {author} {\bibfnamefont {Ian}\ \bibnamefont
  {Fischer}}, \ and\ \bibinfo {author} {\bibfnamefont {Joshua~V}\ \bibnamefont
  {Dillon}},\ }\bibfield  {title} {\enquote {\bibinfo {title} {Uncertainty in
  the variational information bottleneck},}\ }\href@noop {} {\bibfield
  {journal} {\bibinfo  {journal} {arXiv preprint arXiv:1807.00906}\ } (\bibinfo
  {year} {2018}{\natexlab{a}})}\BibitemShut {NoStop}%
\bibitem [{\citenamefont {Amjad}\ and\ \citenamefont
  {Geiger}(2018)}]{amjad2018not}%
  \BibitemOpen
  \bibfield  {author} {\bibinfo {author} {\bibfnamefont {Rana~Ali}\
  \bibnamefont {Amjad}}\ and\ \bibinfo {author} {\bibfnamefont {Bernhard~C}\
  \bibnamefont {Geiger}},\ }\bibfield  {title} {\enquote {\bibinfo {title}
  {Learning representations for neural network-based classification using the
  information bottleneck principle},}\ }\href@noop {} {\bibfield  {journal}
  {\bibinfo  {journal} {arXiv preprint arXiv:1802.09766}\ } (\bibinfo {year}
  {2018})}\BibitemShut {NoStop}%
\bibitem [{\citenamefont {Shwartz-Ziv}\ and\ \citenamefont
  {Tishby}(2017)}]{shwartz2017opening}%
  \BibitemOpen
  \bibfield  {author} {\bibinfo {author} {\bibfnamefont {Ravid}\ \bibnamefont
  {Shwartz-Ziv}}\ and\ \bibinfo {author} {\bibfnamefont {Naftali}\ \bibnamefont
  {Tishby}},\ }\bibfield  {title} {\enquote {\bibinfo {title} {Opening the
  black box of deep neural networks via information},}\ }\href@noop {}
  {\bibfield  {journal} {\bibinfo  {journal} {arXiv preprint arXiv:1703.00810}\
  } (\bibinfo {year} {2017})}\BibitemShut {NoStop}%
\bibitem [{\citenamefont {Saxe}\ \emph {et~al.}(2018)\citenamefont {Saxe},
  \citenamefont {Bansal}, \citenamefont {Dapello}, \citenamefont {Advani},
  \citenamefont {Kolchinsky}, \citenamefont {Tracey},\ and\ \citenamefont
  {Cox}}]{saxe2018information}%
  \BibitemOpen
  \bibfield  {author} {\bibinfo {author} {\bibfnamefont {AM}~\bibnamefont
  {Saxe}}, \bibinfo {author} {\bibfnamefont {Y}~\bibnamefont {Bansal}},
  \bibinfo {author} {\bibfnamefont {J}~\bibnamefont {Dapello}}, \bibinfo
  {author} {\bibfnamefont {M}~\bibnamefont {Advani}}, \bibinfo {author}
  {\bibfnamefont {A}~\bibnamefont {Kolchinsky}}, \bibinfo {author}
  {\bibfnamefont {BD}~\bibnamefont {Tracey}}, \ and\ \bibinfo {author}
  {\bibfnamefont {DD}~\bibnamefont {Cox}},\ }\bibfield  {title} {\enquote
  {\bibinfo {title} {On the information bottleneck theory of deep learning},}\
  }in\ \href@noop {} {\emph {\bibinfo {booktitle} {International Conference on
  Learning Representations}}}\ (\bibinfo {year} {2018})\BibitemShut {NoStop}%
\bibitem [{\citenamefont {Lemar{\'e}chal}(2001)}]{lemarechal2001lagrangian}%
  \BibitemOpen
  \bibfield  {author} {\bibinfo {author} {\bibfnamefont {Claude}\ \bibnamefont
  {Lemar{\'e}chal}},\ }\bibfield  {title} {\enquote {\bibinfo {title}
  {Lagrangian relaxation},}\ }in\ \href@noop {} {\emph {\bibinfo {booktitle}
  {Computational combinatorial optimization}}}\ (\bibinfo  {publisher}
  {Springer},\ \bibinfo {year} {2001})\ pp.\ \bibinfo {pages}
  {112--156}\BibitemShut {NoStop}%
\bibitem [{\citenamefont {Chechik}\ \emph {et~al.}(2005)\citenamefont
  {Chechik}, \citenamefont {Globerson}, \citenamefont {Tishby},\ and\
  \citenamefont {Weiss}}]{chechik_information_2005}%
  \BibitemOpen
  \bibfield  {author} {\bibinfo {author} {\bibfnamefont {Gal}\ \bibnamefont
  {Chechik}}, \bibinfo {author} {\bibfnamefont {Amir}\ \bibnamefont
  {Globerson}}, \bibinfo {author} {\bibfnamefont {Naftali}\ \bibnamefont
  {Tishby}}, \ and\ \bibinfo {author} {\bibfnamefont {Yair}\ \bibnamefont
  {Weiss}},\ }\bibfield  {title} {\enquote {\bibinfo {title} {Information
  bottleneck for {Gaussian} variables},}\ }\href@noop {} {\bibfield  {journal}
  {\bibinfo  {journal} {Journal of Machine Learning Research}\ }\textbf
  {\bibinfo {volume} {6}},\ \bibinfo {pages} {165--188} (\bibinfo {year}
  {2005})}\BibitemShut {NoStop}%
\bibitem [{\citenamefont {Kolchinsky}\ \emph {et~al.}(2018)\citenamefont
  {Kolchinsky}, \citenamefont {Tracey},\ and\ \citenamefont
  {Van~Kuyk}}]{kolchinsky2018caveats}%
  \BibitemOpen
  \bibfield  {author} {\bibinfo {author} {\bibfnamefont {Artemy}\ \bibnamefont
  {Kolchinsky}}, \bibinfo {author} {\bibfnamefont {Brendan~D}\ \bibnamefont
  {Tracey}}, \ and\ \bibinfo {author} {\bibfnamefont {Steven}\ \bibnamefont
  {Van~Kuyk}},\ }\bibfield  {title} {\enquote {\bibinfo {title} {Caveats for
  information bottleneck in deterministic scenarios},}\ }in\ \href@noop {}
  {\emph {\bibinfo {booktitle} {The International Conference on Learning
  Representations (ICLR)}}}\ (\bibinfo {year} {2018})\BibitemShut {NoStop}%
\bibitem [{\citenamefont {Miettinen}(1998)}]{miettinen_nonlinear_1998}%
  \BibitemOpen
  \bibfield  {author} {\bibinfo {author} {\bibfnamefont {Kaisa}\ \bibnamefont
  {Miettinen}},\ }\href {\doibase 10.1007/978-1-4615-5563-6} {\emph {\bibinfo
  {title} {Nonlinear {Multiobjective} {Optimization}}}},\ edited by\ \bibinfo
  {editor} {\bibfnamefont {Frederick~S.}\ \bibnamefont {Hillier}},\ \bibinfo
  {series} {International {Series} in {Operations} {Research} \& {Management}
  {Science}}, Vol.~\bibinfo {volume} {12}\ (\bibinfo  {publisher} {Springer
  US},\ \bibinfo {address} {Boston, MA},\ \bibinfo {year} {1998})\BibitemShut
  {NoStop}%
\bibitem [{\citenamefont {Rodr{\'i}guez~G{\'a}lvez}\ \emph
  {et~al.}(2019)\citenamefont {Rodr{\'i}guez~G{\'a}lvez}, \citenamefont
  {Thobaben},\ and\ \citenamefont {Skoglund}}]{anonymous2020the}%
  \BibitemOpen
  \bibfield  {author} {\bibinfo {author} {\bibfnamefont {Borja}\ \bibnamefont
  {Rodr{\'i}guez~G{\'a}lvez}}, \bibinfo {author} {\bibfnamefont {Ragnar}\
  \bibnamefont {Thobaben}}, \ and\ \bibinfo {author} {\bibfnamefont {Mikael}\
  \bibnamefont {Skoglund}},\ }\bibfield  {title} {\enquote {\bibinfo {title}
  {The convex information bottleneck lagrangian},}\ }\bibfield  {booktitle}
  {\emph {\bibinfo {booktitle} {Submitted to International Conference on
  Learning Representations}},\ }\href@noop {} {\bibfield  {journal} {\bibinfo
  {journal} {arXiv preprint arXiv:1911.11000}\ } (\bibinfo {year}
  {2019})}\BibitemShut {NoStop}%
\bibitem [{\citenamefont {Kolchinsky}\ and\ \citenamefont
  {Tracey}(2017)}]{kolchinsky_upperbound_2017}%
  \BibitemOpen
  \bibfield  {author} {\bibinfo {author} {\bibfnamefont {Artemy}\ \bibnamefont
  {Kolchinsky}}\ and\ \bibinfo {author} {\bibfnamefont {Brendan~D.}\
  \bibnamefont {Tracey}},\ }\bibfield  {title} {\enquote {\bibinfo {title}
  {Estimating mixture entropy with pairwise distances},}\ }\href {\doibase
  10.3390/e19070361} {\bibfield  {journal} {\bibinfo  {journal} {Entropy}\
  }\textbf {\bibinfo {volume} {19}} (\bibinfo {year} {2017}),\
  10.3390/e19070361}\BibitemShut {NoStop}%
\bibitem [{\citenamefont {Chalk}\ \emph {et~al.}(2016)\citenamefont {Chalk},
  \citenamefont {Marre},\ and\ \citenamefont {Tkacik}}]{chalk_relevant_2016}%
  \BibitemOpen
  \bibfield  {author} {\bibinfo {author} {\bibfnamefont {Matthew}\ \bibnamefont
  {Chalk}}, \bibinfo {author} {\bibfnamefont {Olivier}\ \bibnamefont {Marre}},
  \ and\ \bibinfo {author} {\bibfnamefont {Gasper}\ \bibnamefont {Tkacik}},\
  }\bibfield  {title} {\enquote {\bibinfo {title} {Relevant sparse codes with
  variational information bottleneck},}\ }in\ \href@noop {} {\emph {\bibinfo
  {booktitle} {Advances in {Neural} {Information} {Processing} {Systems}}}}\
  (\bibinfo {year} {2016})\ pp.\ \bibinfo {pages} {1957--1965}\BibitemShut
  {NoStop}%
\bibitem [{\citenamefont {Achille}\ and\ \citenamefont
  {Soatto}(2016)}]{achille2016information}%
  \BibitemOpen
  \bibfield  {author} {\bibinfo {author} {\bibfnamefont {Alessandro}\
  \bibnamefont {Achille}}\ and\ \bibinfo {author} {\bibfnamefont {Stefano}\
  \bibnamefont {Soatto}},\ }\bibfield  {title} {\enquote {\bibinfo {title}
  {Information dropout: learning optimal representations through noise},}\
  }\href@noop {} {\bibfield  {journal} {\bibinfo  {journal} {arXiv preprint
  arXiv:1611.01353}\ } (\bibinfo {year} {2016})}\BibitemShut {NoStop}%
\bibitem [{\citenamefont {Goodfellow}\ \emph {et~al.}(2016)\citenamefont
  {Goodfellow}, \citenamefont {Bengio},\ and\ \citenamefont
  {Courville}}]{goodfellow2016deep}%
  \BibitemOpen
  \bibfield  {author} {\bibinfo {author} {\bibfnamefont {Ian}\ \bibnamefont
  {Goodfellow}}, \bibinfo {author} {\bibfnamefont {Yoshua}\ \bibnamefont
  {Bengio}}, \ and\ \bibinfo {author} {\bibfnamefont {Aaron}\ \bibnamefont
  {Courville}},\ }\href@noop {} {\emph {\bibinfo {title} {Deep learning}}}\
  (\bibinfo  {publisher} {MIT Press},\ \bibinfo {year} {2016})\BibitemShut
  {NoStop}%
\bibitem [{\citenamefont {Silverman}(2018)}]{silverman2018density}%
  \BibitemOpen
  \bibfield  {author} {\bibinfo {author} {\bibfnamefont {Bernard~W}\
  \bibnamefont {Silverman}},\ }\href@noop {} {\emph {\bibinfo {title} {Density
  estimation for statistics and data analysis}}}\ (\bibinfo  {publisher}
  {Routledge},\ \bibinfo {year} {2018})\BibitemShut {NoStop}%
\bibitem [{\citenamefont {Kolchinsky}\ and\ \citenamefont
  {Wolpert}(2016)}]{kolchinsky-slip-rnn-2016}%
  \BibitemOpen
  \bibfield  {author} {\bibinfo {author} {\bibfnamefont {Artemy}\ \bibnamefont
  {Kolchinsky}}\ and\ \bibinfo {author} {\bibfnamefont {David~H.}\ \bibnamefont
  {Wolpert}},\ }\bibfield  {title} {\enquote {\bibinfo {title} {Supervised
  learning with information penalties},}\ }in\ \href
  {http://people.idsia.ch/~rupesh/rnnsymposium2016/files/kolchinsky.pdf} {\emph
  {\bibinfo {booktitle} {Recurrent Neural Networks Symposium at NIPS'16}}}\
  (\bibinfo {address} {Barcelona, Spain},\ \bibinfo {year} {2016})\BibitemShut
  {NoStop}%
\bibitem [{\citenamefont {Kolchinsky}\ \emph {et~al.}(2017)\citenamefont
  {Kolchinsky}, \citenamefont {Tracey},\ and\ \citenamefont
  {Wolpert}}]{nonlinearIBv1}%
  \BibitemOpen
  \bibfield  {author} {\bibinfo {author} {\bibfnamefont {Artemy}\ \bibnamefont
  {Kolchinsky}}, \bibinfo {author} {\bibfnamefont {Brendan~D.}\ \bibnamefont
  {Tracey}}, \ and\ \bibinfo {author} {\bibfnamefont {David~H.}\ \bibnamefont
  {Wolpert}},\ }\bibfield  {title} {\enquote {\bibinfo {title} {Nonlinear
  information bottleneck (v1)},}\ }\href {http://arxiv.org/abs/1705.02436v1}
  {\bibfield  {journal} {\bibinfo  {journal} {arXiv e-prints}\ }\textbf
  {\bibinfo {volume} {abs/1705.02436v1}} (\bibinfo {year} {2017})},\ \Eprint
  {http://arxiv.org/abs/1705.02436v1} {arXiv:1705.02436v1} \BibitemShut
  {NoStop}%
\bibitem [{\citenamefont {Schraudolph}(1995)}]{schraudolph_optimization_1995}%
  \BibitemOpen
  \bibfield  {author} {\bibinfo {author} {\bibfnamefont {Nicol~Norbert}\
  \bibnamefont {Schraudolph}},\ }\emph {\bibinfo {title} {Optimization of
  entropy with neural networks}},\ \href@noop {} {Ph.D. thesis},\ \bibinfo
  {school} {Citeseer} (\bibinfo {year} {1995})\BibitemShut {NoStop}%
\bibitem [{\citenamefont
  {Schraudolph}(2004)}]{schraudolph_gradient-based_2004}%
  \BibitemOpen
  \bibfield  {author} {\bibinfo {author} {\bibfnamefont {Nicol~N.}\
  \bibnamefont {Schraudolph}},\ }\bibfield  {title} {\enquote {\bibinfo {title}
  {Gradient-based manipulation of nonparametric entropy estimates},}\
  }\href@noop {} {\bibfield  {journal} {\bibinfo  {journal} {Neural Networks,
  IEEE Transactions on}\ }\textbf {\bibinfo {volume} {15}},\ \bibinfo {pages}
  {828--837} (\bibinfo {year} {2004})}\BibitemShut {NoStop}%
\bibitem [{\citenamefont {Shwartz}\ \emph {et~al.}(2005)\citenamefont
  {Shwartz}, \citenamefont {Zibulevsky},\ and\ \citenamefont
  {Schechner}}]{shwartz_fast_2005}%
  \BibitemOpen
  \bibfield  {author} {\bibinfo {author} {\bibfnamefont {Sarit}\ \bibnamefont
  {Shwartz}}, \bibinfo {author} {\bibfnamefont {Michael}\ \bibnamefont
  {Zibulevsky}}, \ and\ \bibinfo {author} {\bibfnamefont {Yoav~Y.}\
  \bibnamefont {Schechner}},\ }\bibfield  {title} {\enquote {\bibinfo {title}
  {Fast kernel entropy estimation and optimization},}\ }\href@noop {}
  {\bibfield  {journal} {\bibinfo  {journal} {Signal Processing}\ }\textbf
  {\bibinfo {volume} {85}},\ \bibinfo {pages} {1045--1058} (\bibinfo {year}
  {2005})}\BibitemShut {NoStop}%
\bibitem [{\citenamefont {Torkkola}(2003)}]{torkkola2003feature}%
  \BibitemOpen
  \bibfield  {author} {\bibinfo {author} {\bibfnamefont {Kari}\ \bibnamefont
  {Torkkola}},\ }\bibfield  {title} {\enquote {\bibinfo {title} {Feature
  extraction by non-parametric mutual information maximization},}\ }\href@noop
  {} {\bibfield  {journal} {\bibinfo  {journal} {Journal of machine learning
  research}\ }\textbf {\bibinfo {volume} {3}},\ \bibinfo {pages} {1415--1438}
  (\bibinfo {year} {2003})}\BibitemShut {NoStop}%
\bibitem [{\citenamefont {Hlavávcková-Schindler}\ \emph
  {et~al.}(2007)\citenamefont {Hlavávcková-Schindler}, \citenamefont {Palus},
  \citenamefont {Vejmelka},\ and\ \citenamefont
  {Bhattacharya}}]{hlavavckova-schindler_causality_2007}%
  \BibitemOpen
  \bibfield  {author} {\bibinfo {author} {\bibfnamefont {Katerina}\
  \bibnamefont {Hlavávcková-Schindler}}, \bibinfo {author} {\bibfnamefont
  {Milan}\ \bibnamefont {Palus}}, \bibinfo {author} {\bibfnamefont {Martin}\
  \bibnamefont {Vejmelka}}, \ and\ \bibinfo {author} {\bibfnamefont {Joydeep}\
  \bibnamefont {Bhattacharya}},\ }\bibfield  {title} {\enquote {\bibinfo
  {title} {Causality detection based on information-theoretic approaches in
  time series analysis},}\ }\href@noop {} {\bibfield  {journal} {\bibinfo
  {journal} {Physics Reports}\ }\textbf {\bibinfo {volume} {441}},\ \bibinfo
  {pages} {1--46} (\bibinfo {year} {2007})}\BibitemShut {NoStop}%
\bibitem [{\citenamefont {Goodfellow}\ \emph {et~al.}(2014)\citenamefont
  {Goodfellow}, \citenamefont {Pouget-Abadie}, \citenamefont {Mirza},
  \citenamefont {Xu}, \citenamefont {Warde-Farley}, \citenamefont {Ozair},
  \citenamefont {Courville},\ and\ \citenamefont
  {Bengio}}]{goodfellow2014generative}%
  \BibitemOpen
  \bibfield  {author} {\bibinfo {author} {\bibfnamefont {Ian}\ \bibnamefont
  {Goodfellow}}, \bibinfo {author} {\bibfnamefont {Jean}\ \bibnamefont
  {Pouget-Abadie}}, \bibinfo {author} {\bibfnamefont {Mehdi}\ \bibnamefont
  {Mirza}}, \bibinfo {author} {\bibfnamefont {Bing}\ \bibnamefont {Xu}},
  \bibinfo {author} {\bibfnamefont {David}\ \bibnamefont {Warde-Farley}},
  \bibinfo {author} {\bibfnamefont {Sherjil}\ \bibnamefont {Ozair}}, \bibinfo
  {author} {\bibfnamefont {Aaron}\ \bibnamefont {Courville}}, \ and\ \bibinfo
  {author} {\bibfnamefont {Yoshua}\ \bibnamefont {Bengio}},\ }\bibfield
  {title} {\enquote {\bibinfo {title} {Generative adversarial nets},}\ }in\
  \href@noop {} {\emph {\bibinfo {booktitle} {Advances in neural information
  processing systems}}}\ (\bibinfo {year} {2014})\ pp.\ \bibinfo {pages}
  {2672--2680}\BibitemShut {NoStop}%
\bibitem [{\citenamefont {Hinton}\ and\ \citenamefont
  {Zemel}(1994)}]{hinton_autoencoders_1994}%
  \BibitemOpen
  \bibfield  {author} {\bibinfo {author} {\bibfnamefont {Geoffrey~E.}\
  \bibnamefont {Hinton}}\ and\ \bibinfo {author} {\bibfnamefont {Richard~S.}\
  \bibnamefont {Zemel}},\ }\bibfield  {title} {\enquote {\bibinfo {title}
  {Autoencoders, minimum description length, and {Helmholtz} free energy},}\
  }\href@noop {} {\bibfield  {journal} {\bibinfo  {journal} {Advances in neural
  information processing systems}\ ,\ \bibinfo {pages} {3--3}} (\bibinfo {year}
  {1994})}\BibitemShut {NoStop}%
\bibitem [{\citenamefont {Hinton}\ and\ \citenamefont
  {Zemel}(1997)}]{hinton_minimizing_1997}%
  \BibitemOpen
  \bibfield  {author} {\bibinfo {author} {\bibfnamefont {Geoffrey~E.}\
  \bibnamefont {Hinton}}\ and\ \bibinfo {author} {\bibfnamefont {Richard~S.}\
  \bibnamefont {Zemel}},\ }\bibfield  {title} {\enquote {\bibinfo {title}
  {Minimizing description length in an unsupervised neural network},}\
  }\href@noop {} {\bibfield  {journal} {\bibinfo  {journal} {Preprint}\ }
  (\bibinfo {year} {1997})}\BibitemShut {NoStop}%
\bibitem [{\citenamefont {Deco}\ \emph {et~al.}(1993)\citenamefont {Deco},
  \citenamefont {Finnoff},\ and\ \citenamefont
  {Zimmermann}}]{deco_elimination_1993}%
  \BibitemOpen
  \bibfield  {author} {\bibinfo {author} {\bibfnamefont {G.}~\bibnamefont
  {Deco}}, \bibinfo {author} {\bibfnamefont {W.}~\bibnamefont {Finnoff}}, \
  and\ \bibinfo {author} {\bibfnamefont {H.~G.}\ \bibnamefont {Zimmermann}},\
  }\bibfield  {title} {\enquote {\bibinfo {title} {Elimination of
  {Overtraining} by a {Mutual} {Information} {Network}},}\ }in\ \href@noop {}
  {\emph {\bibinfo {booktitle} {ICANN '93}}},\ \bibinfo {editor} {edited by\
  \bibinfo {editor} {\bibfnamefont {Stan}\ \bibnamefont {Gielen}}\ and\
  \bibinfo {editor} {\bibfnamefont {Bert}\ \bibnamefont {Kappen}}}\ (\bibinfo
  {publisher} {Springer London},\ \bibinfo {year} {1993})\ pp.\ \bibinfo
  {pages} {744--749},\ \bibinfo {note} {dOI:
  10.1007/978-1-4471-2063-6\_208}\BibitemShut {NoStop}%
\bibitem [{\citenamefont {Vincent}\ \emph {et~al.}(2008)\citenamefont
  {Vincent}, \citenamefont {Larochelle}, \citenamefont {Bengio},\ and\
  \citenamefont {Manzagol}}]{vincent2008extracting}%
  \BibitemOpen
  \bibfield  {author} {\bibinfo {author} {\bibfnamefont {Pascal}\ \bibnamefont
  {Vincent}}, \bibinfo {author} {\bibfnamefont {Hugo}\ \bibnamefont
  {Larochelle}}, \bibinfo {author} {\bibfnamefont {Yoshua}\ \bibnamefont
  {Bengio}}, \ and\ \bibinfo {author} {\bibfnamefont {Pierre-Antoine}\
  \bibnamefont {Manzagol}},\ }\bibfield  {title} {\enquote {\bibinfo {title}
  {Extracting and composing robust features with denoising autoencoders},}\
  }in\ \href@noop {} {\emph {\bibinfo {booktitle} {Proceedings of the 25th
  international conference on Machine learning}}}\ (\bibinfo {organization}
  {ACM},\ \bibinfo {year} {2008})\ pp.\ \bibinfo {pages}
  {1096--1103}\BibitemShut {NoStop}%
\bibitem [{\citenamefont {Kingma}\ and\ \citenamefont
  {Welling}(2014)}]{kingma2013auto}%
  \BibitemOpen
  \bibfield  {author} {\bibinfo {author} {\bibfnamefont {Diederik~P}\
  \bibnamefont {Kingma}}\ and\ \bibinfo {author} {\bibfnamefont {Max}\
  \bibnamefont {Welling}},\ }\bibfield  {title} {\enquote {\bibinfo {title}
  {Auto-encoding variational bayes},}\ }in\ \href@noop {} {\emph {\bibinfo
  {booktitle} {The International Conference on Learning Representations
  (ICLR)}}}\ (\bibinfo {year} {2014})\BibitemShut {NoStop}%
\bibitem [{\citenamefont {Higgins}\ \emph {et~al.}(2017)\citenamefont
  {Higgins}, \citenamefont {Matthey}, \citenamefont {Pal}, \citenamefont
  {Burgess}, \citenamefont {Glorot}, \citenamefont {Botvinick}, \citenamefont
  {Mohamed},\ and\ \citenamefont {Lerchner}}]{higgins2017beta}%
  \BibitemOpen
  \bibfield  {author} {\bibinfo {author} {\bibfnamefont {Irina}\ \bibnamefont
  {Higgins}}, \bibinfo {author} {\bibfnamefont {Loic}\ \bibnamefont {Matthey}},
  \bibinfo {author} {\bibfnamefont {Arka}\ \bibnamefont {Pal}}, \bibinfo
  {author} {\bibfnamefont {Christopher}\ \bibnamefont {Burgess}}, \bibinfo
  {author} {\bibfnamefont {Xavier}\ \bibnamefont {Glorot}}, \bibinfo {author}
  {\bibfnamefont {Matthew}\ \bibnamefont {Botvinick}}, \bibinfo {author}
  {\bibfnamefont {Shakir}\ \bibnamefont {Mohamed}}, \ and\ \bibinfo {author}
  {\bibfnamefont {Alexander}\ \bibnamefont {Lerchner}},\ }\bibfield  {title}
  {\enquote {\bibinfo {title} {beta-vae: Learning basic visual concepts with a
  constrained variational framework.}}\ }\href@noop {} {\bibfield  {journal}
  {\bibinfo  {journal} {ICLR}\ }\textbf {\bibinfo {volume} {2}},\ \bibinfo
  {pages} {6} (\bibinfo {year} {2017})}\BibitemShut {NoStop}%
\bibitem [{\citenamefont {Alemi}\ \emph
  {et~al.}(2018{\natexlab{b}})\citenamefont {Alemi}, \citenamefont {Poole},
  \citenamefont {Fischer}, \citenamefont {Dillon}, \citenamefont {Saurous},\
  and\ \citenamefont {Murphy}}]{alemi2017fixing}%
  \BibitemOpen
  \bibfield  {author} {\bibinfo {author} {\bibfnamefont {Alex}\ \bibnamefont
  {Alemi}}, \bibinfo {author} {\bibfnamefont {Ben}\ \bibnamefont {Poole}},
  \bibinfo {author} {\bibfnamefont {Ian}\ \bibnamefont {Fischer}}, \bibinfo
  {author} {\bibfnamefont {Josh}\ \bibnamefont {Dillon}}, \bibinfo {author}
  {\bibfnamefont {Rif~A.}\ \bibnamefont {Saurous}}, \ and\ \bibinfo {author}
  {\bibfnamefont {Kevin}\ \bibnamefont {Murphy}},\ }\bibfield  {title}
  {\enquote {\bibinfo {title} {Fixing a broken elbo},}\ }in\ \href
  {http://proceedings.mlr.press/v80/alemi18a.html} {\emph {\bibinfo {booktitle}
  {Proceedings of the 35th International Conference on Machine Learning}}}\
  (\bibinfo {address} {Stockholmsmässan, Stockholm Sweden},\ \bibinfo {year}
  {2018})\ pp.\ \bibinfo {pages} {159--168}\BibitemShut {NoStop}%
\bibitem [{\citenamefont {Kingma}\ and\ \citenamefont
  {Ba}(2015)}]{kingma2015adam}%
  \BibitemOpen
  \bibfield  {author} {\bibinfo {author} {\bibfnamefont {Diederik}\
  \bibnamefont {Kingma}}\ and\ \bibinfo {author} {\bibfnamefont {Jimmy}\
  \bibnamefont {Ba}},\ }\bibfield  {title} {\enquote {\bibinfo {title} {Adam: A
  method for stochastic optimization},}\ }in\ \href@noop {} {\emph {\bibinfo
  {booktitle} {3rd International Conference for Learning Representations}}}\
  (\bibinfo {year} {2015})\BibitemShut {NoStop}%
\bibitem [{\citenamefont {Xiao}\ \emph {et~al.}(2017)\citenamefont {Xiao},
  \citenamefont {Rasul},\ and\ \citenamefont {Vollgraf}}]{xiao2017fashion}%
  \BibitemOpen
  \bibfield  {author} {\bibinfo {author} {\bibfnamefont {Han}\ \bibnamefont
  {Xiao}}, \bibinfo {author} {\bibfnamefont {Kashif}\ \bibnamefont {Rasul}}, \
  and\ \bibinfo {author} {\bibfnamefont {Roland}\ \bibnamefont {Vollgraf}},\
  }\bibfield  {title} {\enquote {\bibinfo {title} {Fashion-mnist: A novel image
  dataset for benchmarking machine learning algorithms},}\ }\href@noop {}
  {\bibfield  {journal} {\bibinfo  {journal} {arXiv preprint arXiv:1708.07747}\
  } (\bibinfo {year} {2017})}\BibitemShut {NoStop}%
\bibitem [{\citenamefont {Pace}\ and\ \citenamefont
  {Barry}(1997)}]{pace1997sparse}%
  \BibitemOpen
  \bibfield  {author} {\bibinfo {author} {\bibfnamefont {R~Kelley}\
  \bibnamefont {Pace}}\ and\ \bibinfo {author} {\bibfnamefont {Ronald}\
  \bibnamefont {Barry}},\ }\bibfield  {title} {\enquote {\bibinfo {title}
  {Sparse spatial autoregressions},}\ }\href@noop {} {\bibfield  {journal}
  {\bibinfo  {journal} {Statistics \& Probability Letters}\ }\textbf {\bibinfo
  {volume} {33}},\ \bibinfo {pages} {291--297} (\bibinfo {year}
  {1997})}\BibitemShut {NoStop}%
\bibitem [{\citenamefont {Pedregosa}\ \emph {et~al.}(2011)\citenamefont
  {Pedregosa}, \citenamefont {Varoquaux}, \citenamefont {Gramfort},
  \citenamefont {Michel}, \citenamefont {Thirion}, \citenamefont {Grisel},
  \citenamefont {Blondel}, \citenamefont {Prettenhofer}, \citenamefont {Weiss},
  \citenamefont {Dubourg}, \citenamefont {Vanderplas}, \citenamefont {Passos},
  \citenamefont {Cournapeau}, \citenamefont {Brucher}, \citenamefont {Perrot},\
  and\ \citenamefont {Duchesnay}}]{scikit-learn}%
  \BibitemOpen
  \bibfield  {author} {\bibinfo {author} {\bibfnamefont {F.}~\bibnamefont
  {Pedregosa}}, \bibinfo {author} {\bibfnamefont {G.}~\bibnamefont
  {Varoquaux}}, \bibinfo {author} {\bibfnamefont {A.}~\bibnamefont {Gramfort}},
  \bibinfo {author} {\bibfnamefont {V.}~\bibnamefont {Michel}}, \bibinfo
  {author} {\bibfnamefont {B.}~\bibnamefont {Thirion}}, \bibinfo {author}
  {\bibfnamefont {O.}~\bibnamefont {Grisel}}, \bibinfo {author} {\bibfnamefont
  {M.}~\bibnamefont {Blondel}}, \bibinfo {author} {\bibfnamefont
  {P.}~\bibnamefont {Prettenhofer}}, \bibinfo {author} {\bibfnamefont
  {R.}~\bibnamefont {Weiss}}, \bibinfo {author} {\bibfnamefont
  {V.}~\bibnamefont {Dubourg}}, \bibinfo {author} {\bibfnamefont
  {J.}~\bibnamefont {Vanderplas}}, \bibinfo {author} {\bibfnamefont
  {A.}~\bibnamefont {Passos}}, \bibinfo {author} {\bibfnamefont
  {D.}~\bibnamefont {Cournapeau}}, \bibinfo {author} {\bibfnamefont
  {M.}~\bibnamefont {Brucher}}, \bibinfo {author} {\bibfnamefont
  {M.}~\bibnamefont {Perrot}}, \ and\ \bibinfo {author} {\bibfnamefont
  {E.}~\bibnamefont {Duchesnay}},\ }\bibfield  {title} {\enquote {\bibinfo
  {title} {Scikit-learn: Machine learning in {P}ython},}\ }\href@noop {}
  {\bibfield  {journal} {\bibinfo  {journal} {Journal of Machine Learning
  Research}\ }\textbf {\bibinfo {volume} {12}},\ \bibinfo {pages} {2825--2830}
  (\bibinfo {year} {2011})}\BibitemShut {NoStop}%
\end{thebibliography}%
